\documentclass[vecphys,multphys,sechang]{svmult}
\pdfoutput=1

\usepackage{ergebnisse} 

\usepackage[authoryear]{natbib}

\usepackage{makeidx}         
\usepackage{graphicx}        
\usepackage{multicol}        
\usepackage[bottom]{footmisc}

\makeindex             
\let\oldfootsep=\footnotesep
\def\VEV#1{\left\langle #1\right\rangle}
\newcommand\ltsima{$\; \buildrel <\over\sim \;$}
\newcommand\simlt{\lower.5ex\hbox{\ltsima}}
\newcommand\gtsima{$\; \buildrel >\over\sim \;$}
\newcommand\simgt{\lower.5ex\hbox{\gtsima}}
\newcommand\etal{et~al.}

\newcommand\msun {M_\odot}

\newcommand\rsun {R_\odot}

\newcommand\mearth {M_\oplus}

\newcommand\vperp{v_{\rm \perp}}

\newcommand\pac{Paczy{\'n}ski }
\newcommand\ie{{\it i.e. }}

\newcommand\apj{{\it Astrophys. J.}}
\newcommand\apjl{{\it Astrophys. J. Lett.}}
\newcommand\nat{{\it Nature}}
\newcommand\pasp{{\it P.A.S.P.}}
\newcommand\mnras{{\it M.N.R.A.S.}}
\newcommand\aap{{\it Astron. \& Astrophys.}}
\newcommand\aj{{\it Astron. J.}}

\newcommand\iaucirc{{\it I.A.U.C.}}
\newcommand\rep {\tilde{r}_e}
\newcommand{\mathbold}[1]{\mbox{\boldmath $\bf#1$}}

\begin{document}

\title{Detection of Extrasolar Planets by Gravitational Microlensing}
\author{David P. Bennett\\
University of Notre Dame}
\authorrunning{Exoplanets via Microlensing}
%
\maketitle

\begin{abstract}
Gravitational microlensing provides a unique window on the 
properties and prevalence of extrasolar planetary systems because 
of its ability to find low-mass planets at separations of a few AU. The 
early evidence from microlensing indicates that the most common 
type of exoplanet yet detected are the so-called Òsuper-EarthÓ 
planets of $\sim 10$ Earth-masses at a separation of a few AU from 
their host stars. The detection of two such planets indicates that 
roughly one third of stars have such planets in the separation 
range 1.5-4 AU, which is about an order of magnitude larger 
than the prevalence of gas-giant planets at these separations. 
We review the basic physics of the microlensing method, and 
show why this method allows the detection of Earth-mass planets 
at separations of 2-3 AU with ground-based observations. We 
explore the conditions that allow the detection of the planetary 
host stars and allow measurement of planetary orbital parameters. 
Finally, we show that a low-cost, space-based microlensing survey 
can provide a comprehensive statistical census of extrasolar 
planetary systems with sensitivity down to 0.1 Earth-masses 
at separations ranging from 0.5 AU to infinity.
\end{abstract}

\section{Introduction}
\label{sec-intro}
The gravitational microlensing method relies upon chances alignments
between background source stars and foreground stars, which may
host planet systems. These background source stars serve as sources of
light that are used to probe the gravitational field of the foreground stars
and any planets that they might host. The relative motion of the source star and 
lens system allows the light rays from the source to sample different paths
through the gravitational field of the foreground system, and it is changing
total gravitational lens magnification of the source star with time 
that provides the observable gravitational microlensing signal. 

The microlensing method is unique among exoplanet detection methods
in a number of respects:
\begin{enumerate}
  \item The amplitude of planetary microlensing signals is large
            (typically $\simgt 10$\%) and is approximately
            independent of the planetary mass. Instead, the source-lens
            alignment necessary to give a detectable planetary signal
            depends on the planet-star mass ratio, $q$, and so the 
            probability of a detectable planetary signal scales as $\sim q$.
  \item This scaling of the probability of planet detection with the mass
            ratio, $q$, is shallower than the sensitivity curves for other methods,
            so microlensing is more sensitive to low-mass planets than other
            methods that are sensitive to planetary mass. The sensitivity of
            the microlensing planet search method extends down to $0.1\mearth$.
  \item Microlensing is most sensitive to planets at orbital separations of
            1.5-$4\,$AU, which corresponds to the vicinity of the Einstein ring
            radius. This range of separations also corresponds to the ``snow
            line" where planet formation is most efficient according to the
            leading core accretion model of planet formation. Thus, microlensing
            complements the Doppler radial velocity and transit methods, which
            are most sensitive to planets in very short period orbits.
  \item Microlensing is the only planet detection method that is sensitive
            to old, free-floating planets, which have been ejected from the gravitational
            potential well of their parent stars through planet-planet scattering.
            Theory predicts that such planets may be quite common, and
            ground-based microlensing can detect free-floating gas giant
            planets, while a space-based survey is needed to detect free-floating
            terrestrial planets.
  \item Since the microlensing method doesn't rely upon light from the host star
            in order to detect its planets, it can detect planets orbiting unseen stars.
            This can make it difficult to determine the properties of the host stars, but
            space-based follow-up observations can detect the host stars for
            most planets discovered by microlensing.
  \item A space-based microlensing survey would provide a nearly complete
            statistical census of extrasolar planets with masses down to $0.1\mearth$
            at all separations $\geq 0.5\,$AU. This includes analogs of all the
            Solar System's planets, except for Mercury.
\end{enumerate}

Gravitational microlensing differs from other extrasolar planet search
techniques in a number of aspects. It is a purely gravitational method
that doesn't rely upon on detecting photons from either the planet or
its host star

While most of the known exoplanets have been discovered with
the Doppler radial velocity method, the early results from the
microlensing method indicate that cool, super-Earth or
sub-Neptune mass planets are more representative of 
typical extrasolar planets than any of the $200+$ exoplanets
discovered by radial velocities. 

\section{Gravitational Microlensing Theory}
\label{sec-theory}
\subsection{The Single Lens Case}
\label{sec-singlelens}
\begin{figure}
\centering
\includegraphics[height=5cm]{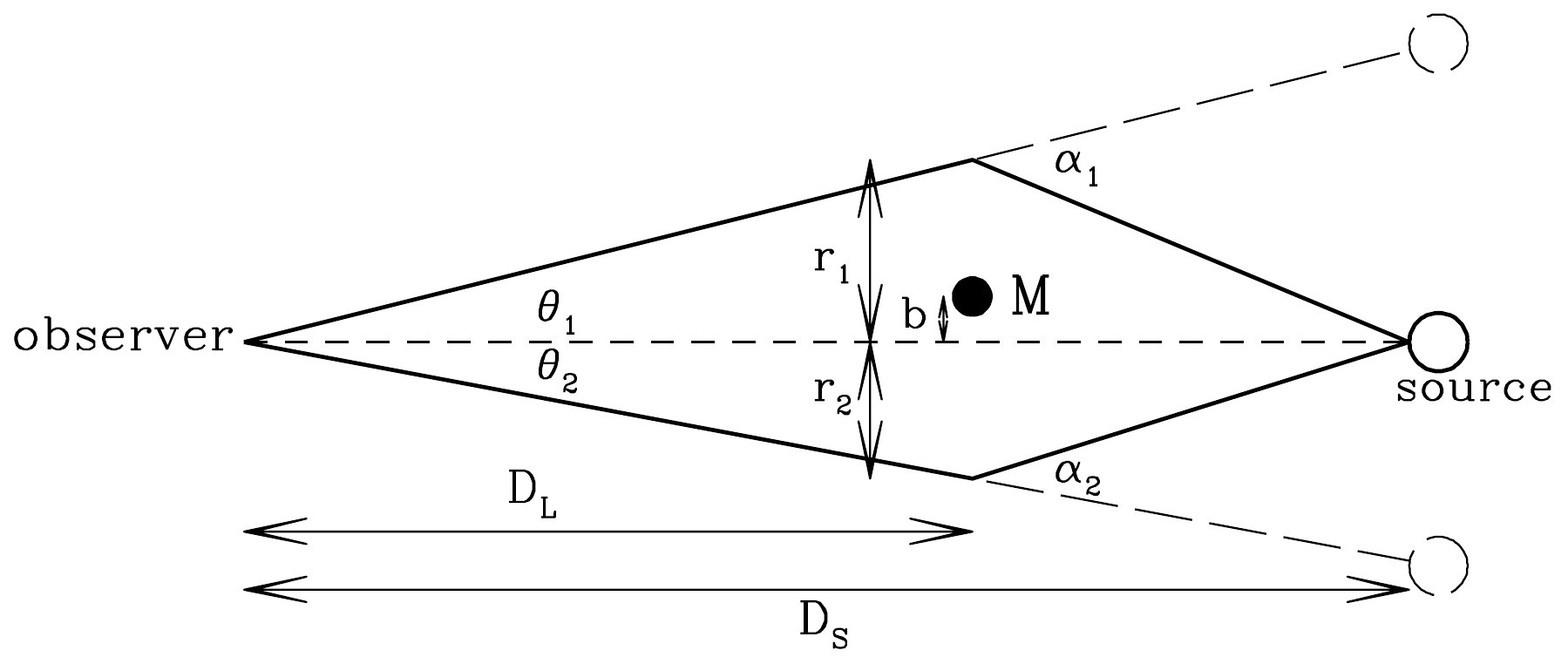}
%
%
\caption{The geometry of gravitational lens of mass, $M$, that is offset
by a distance, $b$, from the line of sight to the source. The observer sees
two images that are offset by angles, $\theta_1$ and $\theta_2$ from the
line of sight to the source star.}
\label{fig-lens_dia}       
\end{figure}
The basic physics of gravitational lensing depends only a single input
from General Relativity, the deflection angle, $\alpha$, for a light ray
passing a mass, $M$, with an impact parameter, $r$:
\begin{equation}
  \alpha = {4GM\over c^2 r} \ .
\label{eq-bend}
\end{equation}
With the lens geometry shown in Fig.~\ref{fig-lens_dia}, we have
\begin{equation}
  \alpha_i = {4GM\over c^2 (r_i - b)} = {r_i D_S\over D_L (D_S-D_L)}  \ ,
\label{eq-bend2}
\end{equation}
in the small angle approximation. If the lens and source are perfectly
aligned, the two images merge to form a ring of radius
\begin{equation}
R_E \equiv \theta_E D_L \equiv \sqrt{4GM D_L(D_S-D_L)\over c^2 \ D_S } \ ,
\label{eq-Re}
\end{equation}
known as the Einstein ring radius. ($\theta_E$ is the angular Einstein
radius.) 
We can now rewrite the single lens equation as
\begin{equation}
r_i = {R_E^2 \over r_i - b } \ ,
\label{eq-1lens_eq}
\end{equation}
and it has two solutions: $r_{+,-} = 0.5(b\pm \sqrt{b^2+4R_E^2})$.
The lensed images are also magnified, and the magnification of a
source of infinitesimal size can be computed using area elements
obtained by differentiating eq.~\ref{eq-1lens_eq}. This yields
\begin{equation}
A_{+,-} = {1\over 2} \left( {u^2+2\over u\sqrt{u^2+4} } \pm 1 \right) \ ,
\label{eq-Apm}
\end{equation}
where $u \equiv b/R_E$ is the dimensionless lens-source separation.
The total magnification of both images is given by
\begin{equation}
A = A_+ + A_- = {u^2+2\over u\sqrt{u^2+4} }  \ .
\label{eq-A}
\end{equation}

For a lens of $M = 1\msun$, that is half-way to a source in the
Galactic center (at $D_S = 8\,$kpc), we find
\begin{equation}
R_E = 4.04\,{\rm AU} \sqrt{{M\over \msun} {D_S\over 8\,{\rm kpc} }\  4x(1-x)} \ ,
\label{eq-ReAU}
\end{equation}
where $x = D_L/D_S$, so $R_E$ is similar to the orbital radius
of planets in our own Solar System. This also implies that 
$\theta_E \sim 1\,$mas. Since the image separation is of order
$\sim \theta_E$, this implies that images will not generally be
resolved with virtually all planned and future astronomical
instruments (with a few exceptions \citep{vlti}). On the other hand,
if we assume a typical Galactic velocity of $\vperp = 100\,$km/sec for the
relative velocity between the lens star and the line-of-sight to the
source, then the typical Einstein radius crossing time for a lens in the
Galactic disk and a bulge source is $t_E = R_E/\vperp \sim 2\,$months.
Thus, the main observational effect for lensing by stars within the
Milky Way is the time varying magnification instead of the image
separation, and this is why it is referred to as microlensing instead
of lensing.

\begin{figure}
\centering
\includegraphics[height=5cm]{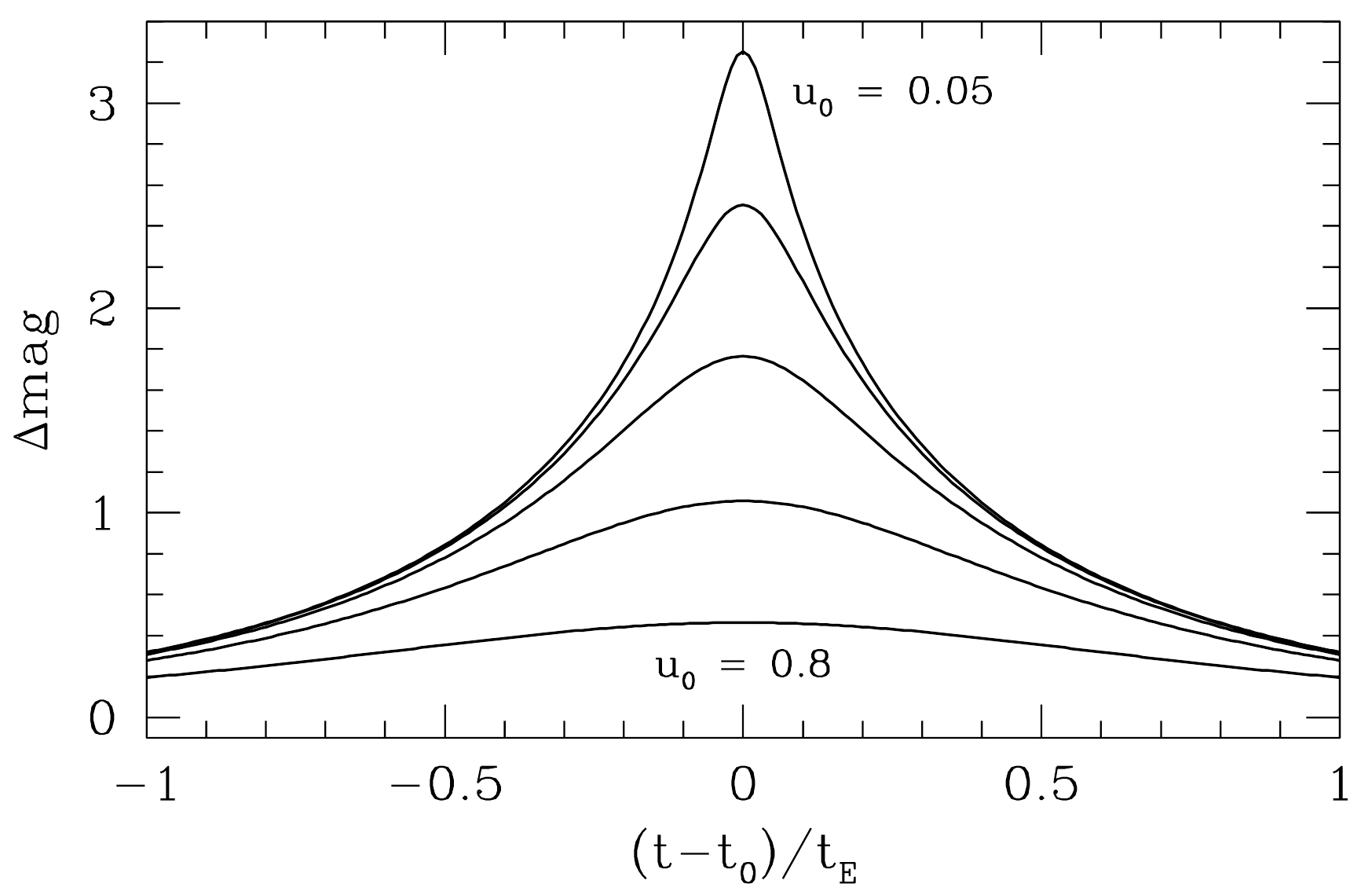}
%
%
\caption{Example microlensing light curves for a point source and a single lens
that moves with a constant lens velocity with respect to the observer-source 
line-of-sight. Light curves with $u_0 = 0.05, 0.1, 0.2, 0.4,$ and 0.8 are
shown.}
\label{fig-lcex}       
\end{figure}

The microlensing light curve is generally described by eq.~\ref{eq-A}
with the lens-source separation given by 
\begin{equation}
u = \sqrt{\left(t-t_0\over t_E\right)^2 + u_0^2 } \ ,
\label{eq-u}
\end{equation}
assuming that the relative motion between the lens and the observer-source
line-of-sight. Thus, a single-lens microlensing light curve is described by
three parameters, the time of peak magnification, $t_0$, the Einstein
radius crossing time or width, $t_E$, and the minimum separation,
$u_0$, which determines the peak magnification. $u_0$ is the
only parameter that affects the intrinsic light curve shape, as shown in 
Fig.~\ref{fig-lcex}, but only $t_E$ constraints the physically interesting 
parameters of the event: the lens mass, $M$, the lens distance, $D_L$,
and the relative velocity, $\vperp$.

The first microlensing events were discovered in 1993 towards
the Large Magellanic Cloud (LMC) by the MACHO Project \citep{macho-nat93}
and towards the Galactic bulge by the OGLE Collaboration.
\citep{ogle93}. The early emphasis of microlensing surveys was
the search for dark matter in the Milky Way's halo \citep{pac86},
but this issue has been largely resolved with the demonstration
that the excess microlensing seen toward the LMC by the
MACHO group\citep{macho_lmc5.7,lmc_tau_contam} requires
at most 20\% of the Milky Way's dark matter in the form of
stellar mass objects, while the results of the EROS group
\citep{eros_lmc2007} suggest that much of this microlensing
excess may be caused by stars associated with the LMC
itself \citep{sahu94}, perhaps in the LMC halo \citep{wu94}.

With the dark matter microlensing question mostly resolved,
the prime focus of microlensing observations has shifted to the 
detection of extrasolar planets. Microlensing was first suggested
as a method to find planets by \citet{liebes}, but as \citet{mao-pac}
pointed out, this requires a consideration of multiple lens
systems.

\subsection{Multiple Lens Systems}
\label{sec-multilens}

The lens equation for a multiple lens system is a straight forward 
generalization of the single lens equation, eq.~\ref{eq-1lens_eq},
but with more than one lens mass, we can no longer assume
that the source, observer and lens system all lie in a single plane.
However, as long as the distances to the lens and source ($D_L$
and $D_S$) are much larger than the extent of the lens system,
we can assume that the lens system resides a single distance,
and define the Einstein radius of the total lens system mass
using eq.~\ref{eq-Re}. So, as before, we will rescale all the length
variables with $R_E$.

Because we can no longer define a source-lens-observer plane,
we must now define the lens and source positions in the 2-dimensional
``lens-plane" perpendicular to the line-of-sight and projected to the
distance of the lens (or equivalently, we can just use angular
variables for the positions of the source and lenses on the sky).
The double-lens system was first solved using two real coordinates
for the lens plane \citep{schneider_weiss}, but the algebra is much 
simpler if we describe the lens plane with complex coordinates
following \citet{bkn73,witt90} and \citet{rhie97}. The generalization of
eq.~\ref{eq-1lens_eq} is
\begin{equation}
w = z - \sum_i {\epsilon_i \over \bar{z} - \bar{x}_i} \ ,
\label{eq-mult_lens}
\end{equation}
where $w$ and $z$ are the complex positions of the source and
image, respectively, and $x_i$ are the complex positions of the 
lens masses. The individual lens masses are represented by
$\epsilon_i$, which is the mass fraction of the $i$th lens mass,
so that $\sum_i \epsilon_i = 1$. The appearance of the complex
conjugates in the denominator in the sum on the right side of 
eq.~\ref{eq-mult_lens} is simply a reflection of the fact that the
lens deflection is the in the direction from the source to the lens
with a magnitude of the inverse of that distance. With real
coordinates, we would express this as the vector difference
of the positions divided by this vector squared, but with complex
coordinates, we can divide through by this vector leaving only
its complex conjugate in the denominator.

If we knew the position of the images, $z$, in eq.~\ref{eq-mult_lens}, then
it would be trivial to solve for the position of the source. But this
is the inverse of the problem that we will usually want to solve, which
is to find the positions of the images based on a
known position for the source. However, the solution of this
``inverse" problem is the basis of the brute-force, ray-shooting
method \citep{schneider_weiss87} for solving eq.~\ref{eq-mult_lens}.
This method involves taking a large grid of points in the ``image
plane" and propagating them back to the source plane using
eq.~\ref{eq-mult_lens}. This method has the advantage that
it can handle very complicated lens mass distributions, but
it is usually not the method of choice for the analysis of
microlensing events.

The most successful method for calculating multi-lens microlensing
light curves \citep{em_planet} involves solving 
eq.~\ref{eq-mult_lens} for the positions
of the point-source images and invoking the ray-shooting method
only in the vicinity of images that are affected by finite-source 
size effects. For the majority of the light curve, the finite-source calculations
are not needed, and we can use the point source magnification
formula. This formula can be derived from the Jacobian 
determinant of the lens equation (and its complex conjugate):
\begin{equation}
J = {\partial w\over \partial z}  {\partial \bar{w}\over \partial \bar{z}} 
  -  {\partial w\over \partial \bar{z}} {\partial \bar{w}\over \partial  z}
  = 1 - \left| {\partial w\over \partial \bar{z}} \right|^2 \ ,
\label{eq-J}
\end{equation}
where 
\begin{equation}
 {\partial w\over \partial \bar{z}}
   =  \sum_i {\epsilon_i \over (\bar{z} - \bar{x}_i)^2} \ .
\label{eq-partw}
\end{equation}
Because eq.~\ref{eq-J} gives the Jacobian determinant of the 
inverse mapping from the image plane to the source plane,
the magnification of each image is given by
\begin{equation}
 A = {1\over |J|} \ ,
\label{eq-AJ}
\end{equation}
evaluated at the position of each image.

The solution of the lens equation, \ref{eq-mult_lens}, is non-trivial.
For the case of two lens masses, this equation can be embedded 
into a fifth order polynomial equation in $z$, which can be solved
numerically. This equation has either 3 or 5 solutions \citep{witt90, rhie97} 
that correspond
to solutions of  \ref{eq-mult_lens}, which means that a double lens
system must have either 3 or 5 images depending on the configuration
of the lens system and the location of the source. For the triple lens
case (which is relevant for at least one planetary microlensing event),
the lens equation can be embedded in a rather complicated tenth
order polynomial that has 4, 6, 8, or 10 solutions that correspond
to physical images \citep{rhie_triple}. This tenth order polynomial
equation can be solved numerically, although it may require
extended precision numerical calculations in order to avoid
serious round-off errors (Bennett et al., in preparation). The
case of 4 lens masses, has also been investigated \citep{rhie_quad},
but the lens equation has not been converted to a polynomial.

The most important feature of lensing by multiple masses occurs
at the locations where $J=0$. From eq.~\ref{eq-AJ}, this implies
infinite magnification for a point source. (The magnification
is always finite for the realistic case of a source of finite angular size.)
For a single lens, $J=0$ only occurs at a single point in the source
plane, the location of the lens mass, but for lens systems with more
than one mass, there are a set of one or more closed curves with $J=0$,
known as critical curves. The source positions corresponding to the
critical curves are obtained by applying the lens equation,
\ref{eq-mult_lens}, and they are referred to as caustic curves.
When the source passes to the interior of a caustic curve, two
new images are created, and it is these new images that
have infinite magnification for the (unphysical) case of
a point source. The shape of the light curve of a (point) source
crossing a caustic has a characteristic form:
\begin{equation}
A = {F_c \Theta(x-x_c) \over \sqrt{x-x_c}} + A_{nc} \ ,
\label{eq-fold_caustic}
\end{equation}
where $F_c$ gives the amplitude of the caustic, $A_{nc}$
gives the magnification of the images that are not associated
with the caustic and $\Theta(x) = 1$ for
$x \geq 0$ and $\Theta(x) = 0$ otherwise. $x$ is the distance 
perpendicular to the direction of the caustic curve, and $x_c$
is the location of the caustic curve. Eq.~\ref{eq-fold_caustic} is
a good approximation to the magnification for a point source
when the curvature of the caustic curve can be neglected.
Note, that the singularity in eq.~\ref{eq-fold_caustic} is weak
enough so that the integral of this formula will yield a
finite magnification for a finite size source star.

Caustic crossings that follow the form of eq.~\ref{eq-fold_caustic}
are often referred to as {\it fold} caustic crossings, and they
have the feature that there is essentially no warning that
the caustic crossing is imminent when the caustic curve is
approached from the outside (\ie $x < x_c$). This is
because the magnification pattern for a fold caustic extends
only to the interior of the caustic since it involves the
magnification of images that only exist inside the caustic
curve. However, each caustic curve also has at least 
three sharp pointy features, known as cusps, and the magnification
pattern extends outward from the cusps on a caustic curve.
The magnification scales as the inverse of the distance to
the cusp, just as in the single lens case, eq.~\ref{eq-A}.

The path of the source with respect to the caustic curves
provides the basic characteristics of a multiple lens 
microlensing light curve. Multiple lens light curves frequently
have features which match the expected $A\sim \Theta (x) x^{-1/2}$
shape of a caustic crossing or the $A\sim r^{-1}$ shape of 
a cusp approach. But, there are additional complications,
as the strength of a caustic crossing ($F_c$ in eq.~\ref{eq-fold_caustic})
can vary and the angular size of the source star can sometimes
be larger than the entire caustic curve for a planetary microlensing
event.

\section{Planetary Microlensing Events}
\label{sec-planet_ev}

Planetary microlensing events are a subset of multiple lens events
where the mass ratio is quite small. Planetary events have light curves
that appear quite similar to the single lens light curves shown in 
Fig.~\ref{fig-lcex}, but for a brief period of time they deviate from the
single lens form and display the characteristics of a binary lens
light curve. We will define a requirement
on the mass ratio $q \equiv \epsilon_2/\epsilon_1 < 0.03$ to separate
planetary microlensing events from stellar binary events following
\citet{bond-moa53}, because $q\approx 0.03$ is the approximate
location of the ``brown dwarf desert" that appears to separate 
stellar from planetary secondaries. We will also initially only
consider events with only one detectable planet, as these
represent the majority of planetary microlensing events and this
will simplify the discussion.

\begin{figure}
\centering
\includegraphics[height=6.3cm]{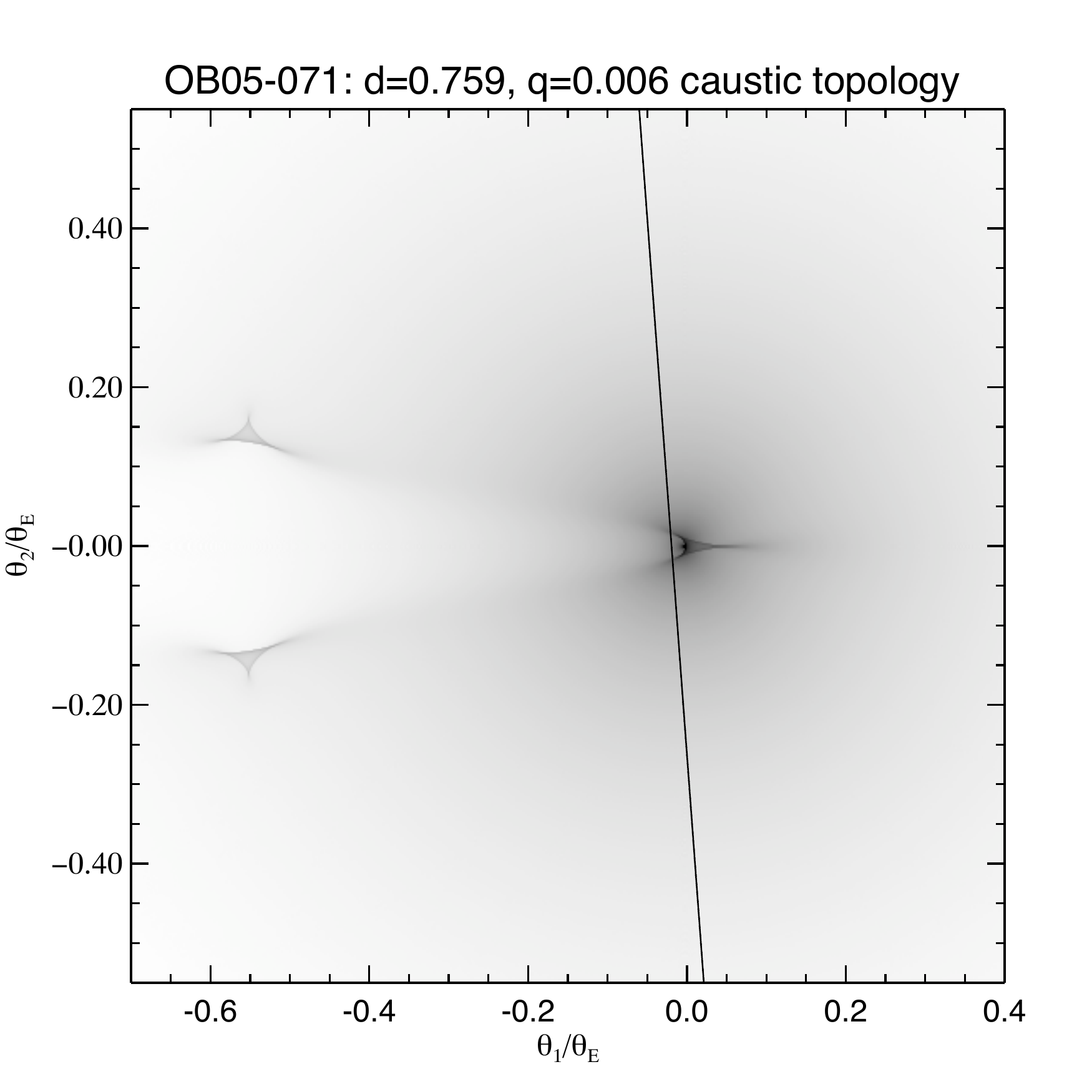}
\includegraphics[height=6.3cm]{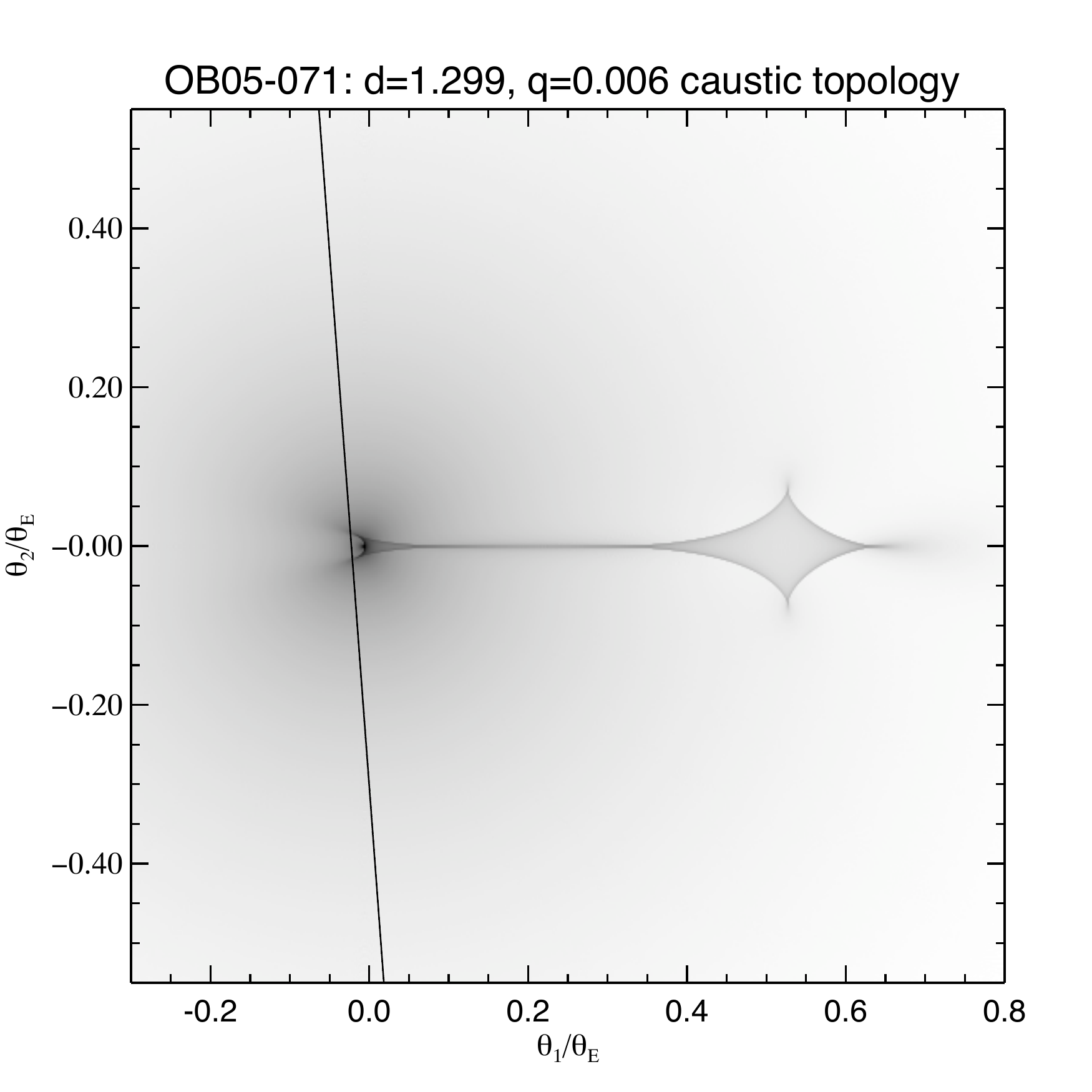}
%
%
\caption{Grey-scale magnification patterns for two models of the
planetary microlensing event OGLE-2005-BLG-71. The darkness
of the image is proportional to the logarithm of the magnification.
The nearly vertical lines in each panel indicate the path of the
source star for each model. The source passes close to the
central, or stellar, caustic, and as discussed in \S~\ref{sec-params},
the magnification pattern
in the vicinity of the central caustic is similar for planetary
systems related by $d\leftrightarrow 1/d$, The magnification
patterns for the planetary caustics (on the outside of each panel)
are clearly very different and easy to distinguish.
(These images are provide courtesy of Daniel Kubas.)}
\label{fig-ob71_caus}       
\end{figure}

The caustic structure of a binary lens system is determined by
the mass ratio and the separation of the lenses \citep{schneider_weiss}.
For a separation $d \ll 1$ (in units of $R_E$), there are three caustics,
two triangular caustics with 3 cusps each and a caustic close to the
center of mass which has 4 cusps. These merge into a single caustic
with 5 cusps at $d\sim 1$, which splits into two caustic curves with
4 cusps each for $d \gg 1$. For small values of the mass ratio, $q$,
the division between these regimes occurs near $d\approx 1$, so 
most events have multiple caustic curves. As shown in 
Fig.~\ref{fig-ob71_caus}.

\subsection{Planetary Caustic Perturbations}
\label{sec-planet_caustic}

There are two classes of caustic curves in planetary microlensing events:
planetary caustics and the stellar or central caustic. The planetary
caustics result when one of the two light rays in Fig.~\ref{fig-lens_dia}
passes close to the planet and is deflected by the planetary gravitational
field of the planet. The locations of the caustics are given by
\begin{equation}
s_c = d - 1/d \ . 
\label{eq-s_c}
\end{equation}
If we see a planetary deviation at a point where the best fit
single lens light curve predicts a magnification, $A$, we can
find the corresponding $u$ value by inverting eq.~\ref{eq-A}
to get $u_c = s_c$,
and then solve for $d$ by inverting eq.~\ref{eq-s_c} to give
\begin{equation}
d = {1\over 2}\left( u_c \pm \sqrt{u_c^2 + 4} \right) \ .
\label{eq-d_sc}
\end{equation}
The two solutions to eq.~\ref{eq-d_sc} for a given $u_c$ value
are referred to as major and minor image perturbations. They
were first studied in detail by \citet{gouldloeb} who showed
that many features of the planetary caustics and their
magnification pattern could be explained by the 
simpler Chang-Refsdal lens system \citep{chang_refs1,chang_refs2}.
Several important features of the planetary caustic light
curve perturbations can be seen in Fig.~\ref{fig-ob71_caus}.
The left side of the left panel of this figure shows the
magnification pattern of the two roughly triangular 
minor image caustic, which are generated by a planet
with $d < 1$. As with all caustics, there is excess magnification
in the interior of the caustic curve, as well as extending outward
from the cusps. But there is also a very pronounced magnification
deficit in between the two minor image caustics, where 
the magnification is substantially below the single lens
magnification. In contrast, the magnification pattern for the
major image caustic (shown on the right side of the
right panel in Fig.~\ref{fig-ob71_caus} is predominantly
positive, with only small magnification deficits very close to
the caustic curve, away from the 4 cusps.

\subsection{Stellar Caustic Perturbations}
\label{sec-star_caustic}

It was originally suspected that the planetary caustic 
perturbations would be the best way to detect planetary signals
in microlensing events, but \citet{griest_saf} argued that 
there were a number of advantages to searching for
planetary light curve perturbations due to the stellar
caustic. They showed that the planet detection efficiency
for each high magnification event was substantially
higher than for events of more modest magnification.
While the higher planet detection efficiency for higher
magnification events was seen in
previous work \citep{bolatto94,em_planet}, \citet{griest_saf}
emphasized that this effect is quite dramatic and that
this fact could be used to increase the observational planet 
detection efficiency. In the same year that the Griest \&
Safizadeh paper was published, the MPS and MOA
Collaborations  demonstrated
this method with observations of the MACHO-98-BLG-35
event. The subsequent analysis showed \citep{mps-98blg35}
that the lens star for this event
did not have any Jupiter-mass planets with a projected
separation of 0.6-8$\,$AU.

The high planet detection efficiency for high magnification events is 
particularly useful when a large number of microlensing events
are discovered by the microlensing survey groups. This is the
current situation, as the OGLE-3 and MOA-2 surveys combine
to detect $>700$ microlensing events in progress toward
the central regions of the Milky Way between February and
October of each year. Relatively
sparse monitoring of events (\ie one or two observations per
day) is required to predict most high magnification events in
advance, and this allows observing resources can be focused on
events with a high planet detection efficiency.

One important consequence of the high planet detection efficiency
for high magnification events is that the chances of detecting
multiple planets in such events are greatly enhanced 
\citep{gaudi_mult}. Indeed, the first multi-planet system
discovered by microlensing is shown below in \S~\ref{sec-obs_detections}.
There is, however, a potential downside to this higher sensitivity to
multiple lens masses. The signals for all the detectable lens masses
will be concentrated in the very high magnification part of the light
curve, and this could make it difficult to work out the details of 
multiple planet systems that are detected in microlensing events.
Thus, the development of efficient light curve modeling methods
for lens systems with three or more masses is an important active
area of current research.

A final advantage of high magnification events is that they allow
planet detection  with relatively faint source stars. This makes it
much easier to detect the planetary host star with follow-up observations
\citep{plan_char} as explained in \S~\ref{sec-theta_E}.

\subsection{Finite Source Effects}
\label{sec-finite_src}

\begin{figure}
\centering
\includegraphics[height=4.8cm]{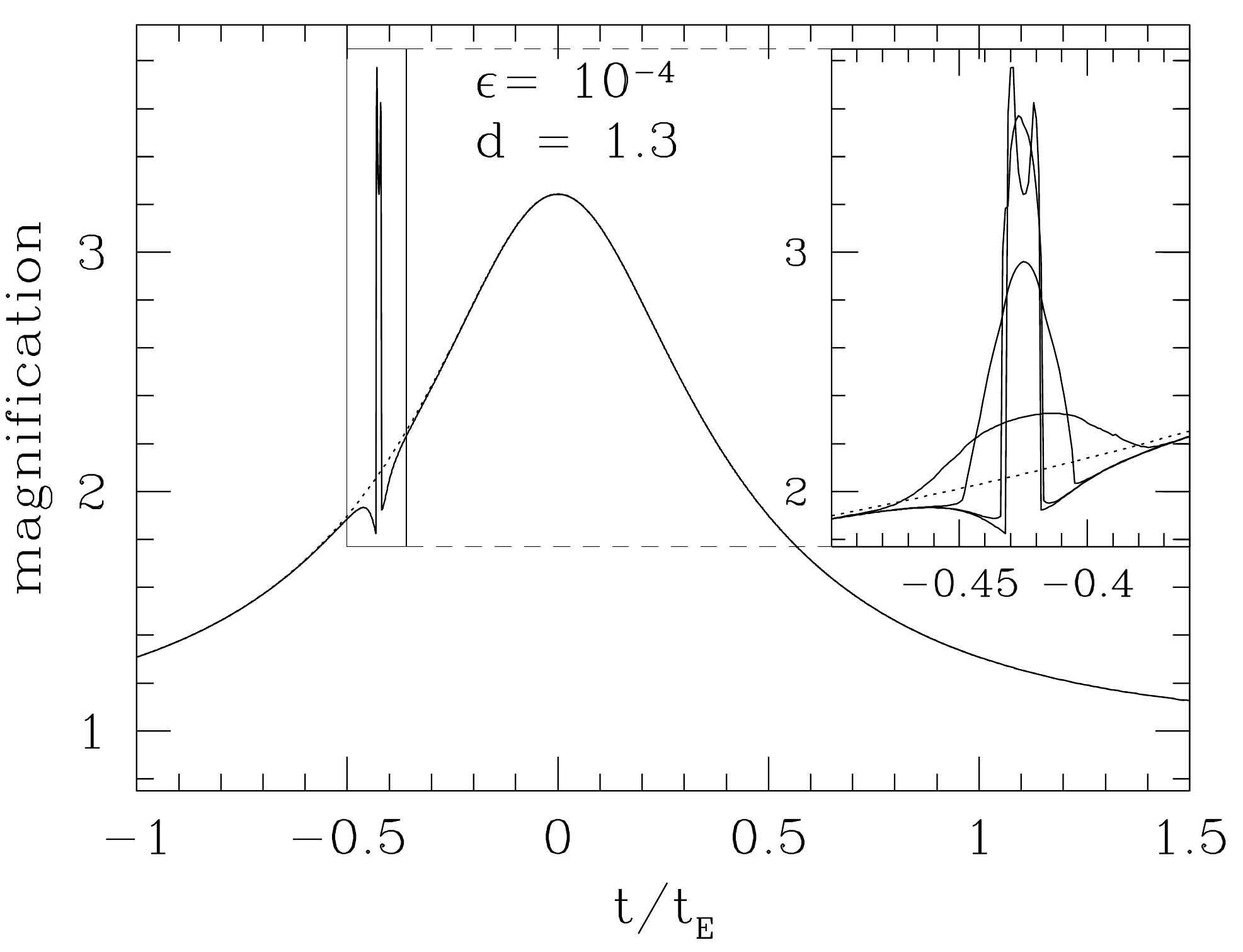}
\includegraphics[height=4.8cm]{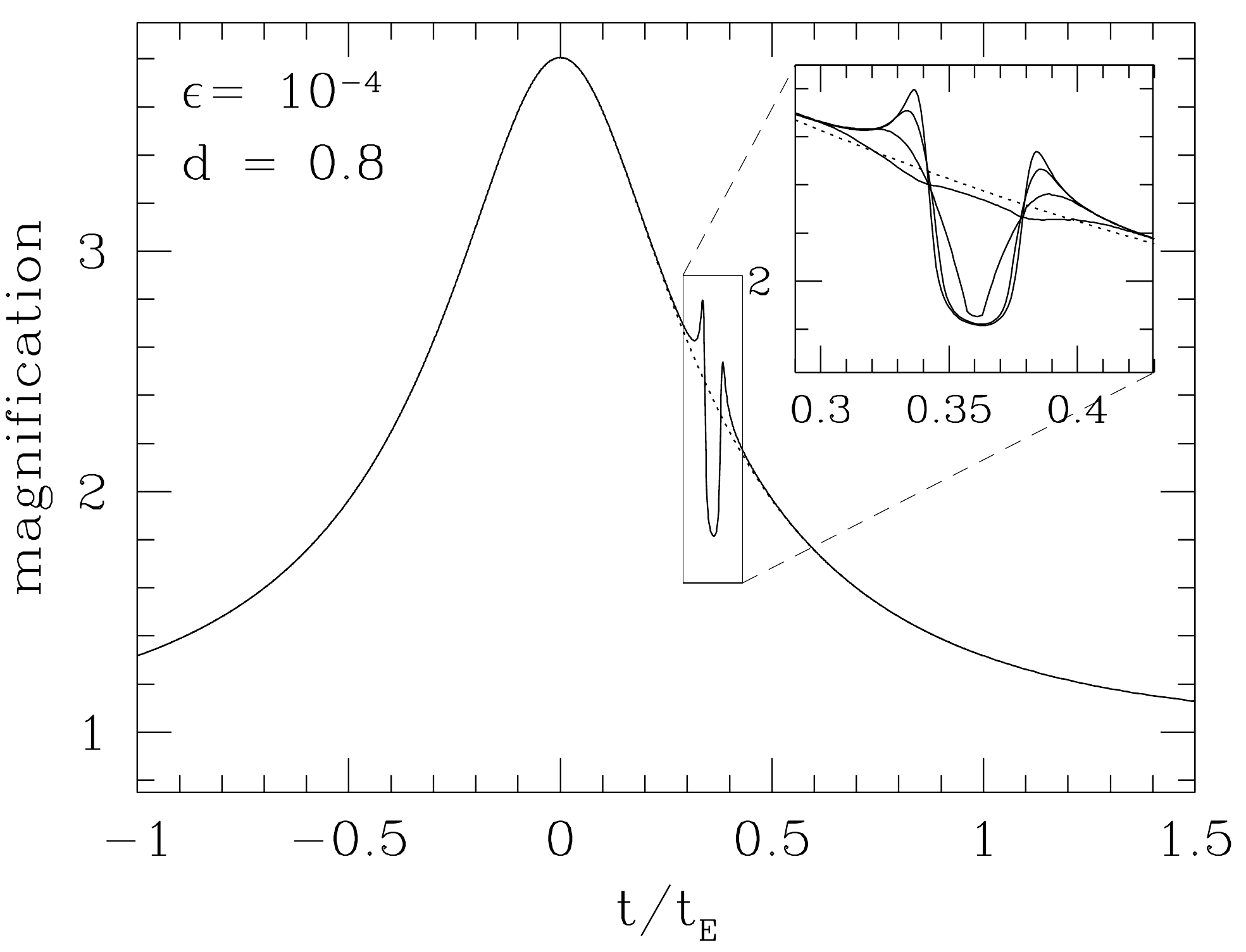}
\includegraphics[height=4.8cm]{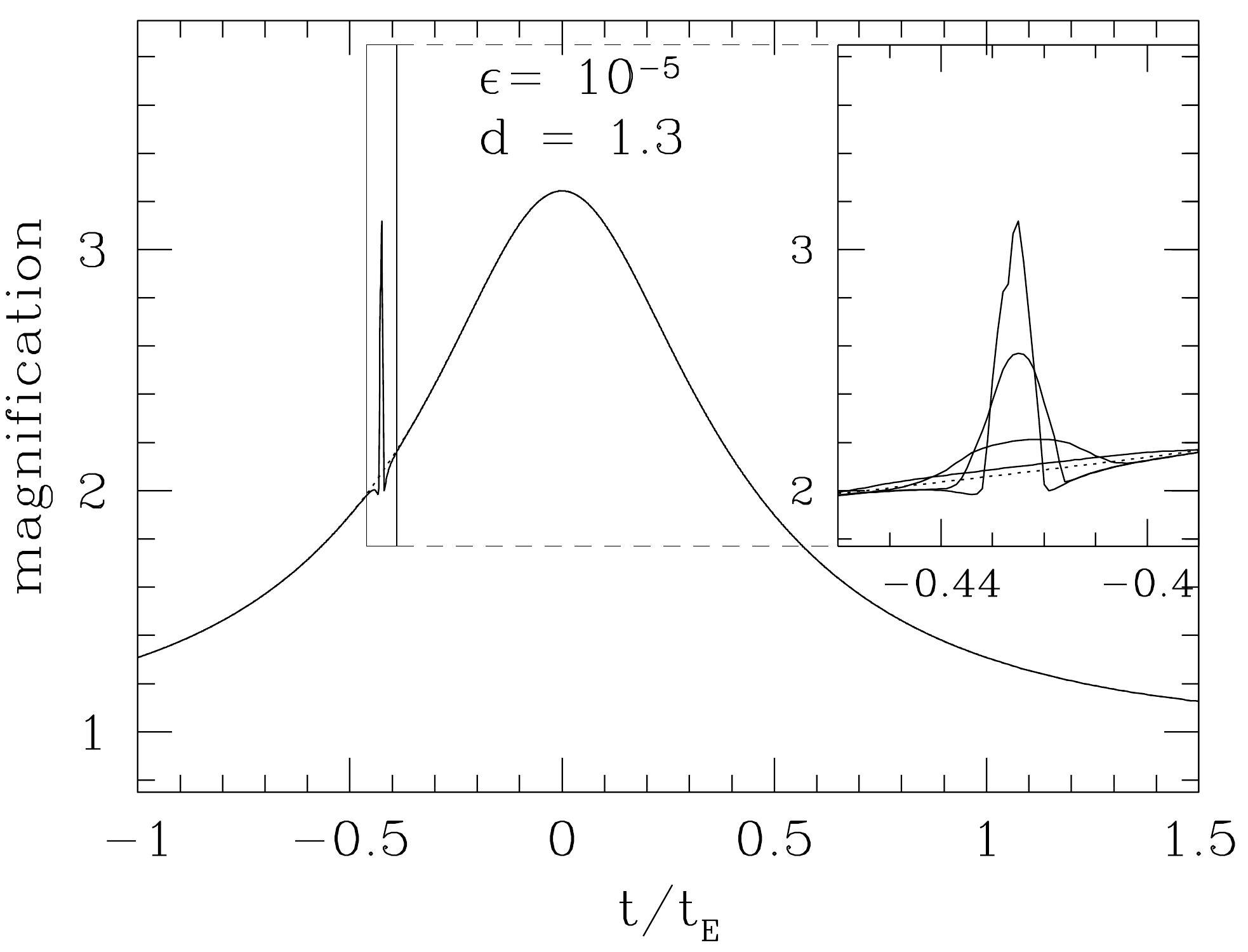}
\includegraphics[height=4.8cm]{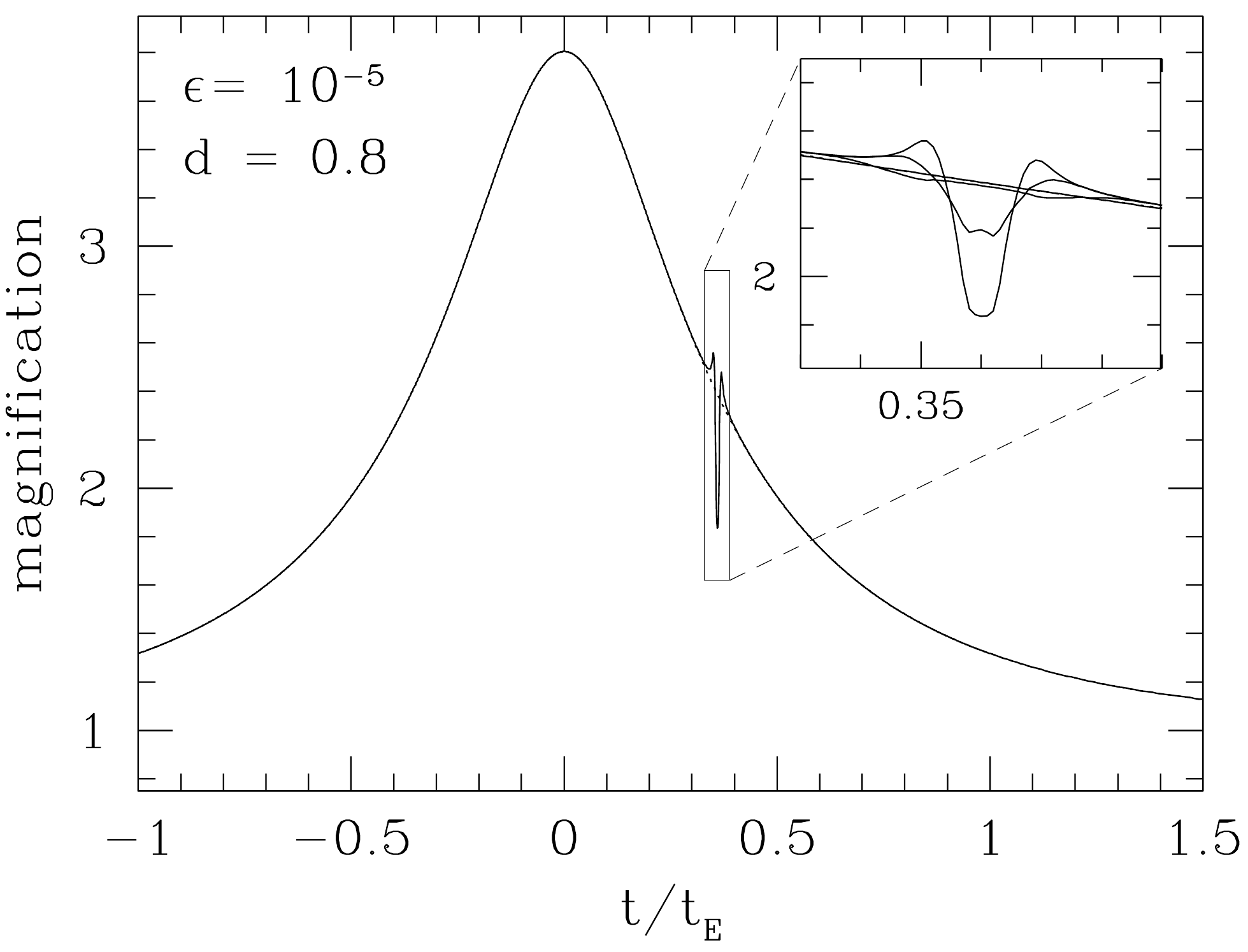}
%
%
\caption{Microlensing lightcurves which show planetary deviations are
  plotted for a mass ratio of $\epsilon = 10^{-4}\ \&\ 10^{-5}$ and separations
  of $d = 1.3\ \&\ 0.8$. The main plots are for a stellar radius of
  $\rho = 0.003$ while the insets show light curves for radii of 0.006, 0.013,
  and 0.03 as well. The dashed curves are the unperturbed single lens
  lightcurves. For each of these lightcurves, the source trajectory
  is at an angle of $\sin^{-1} 0.6$ with respect to the star-planet
  axis. The impact parameter $u_0 = 0.27$ for the $d=0.8$ plots
  and $u_0 = 0.32$ for the $d=1.3$ plots.}
\label{fig-fin_src_lc}       
\end{figure}

Microlensing is arguably the exoplanet search technique that is 
most sensitive to low-mass planets, and lower limit in sensitivity
of the microlensing method is set by the finite angular size of the
source stars. Roughly speaking, when the angular radius
of the planetary Einstein ring, $\sqrt{\epsilon_p} \theta_E$, is
much smaller than the source star angular radius, $\theta_\ast$,
we expect that planetary signal to be washed out. But this is
only a crude, order-of-magnitude estimate, and a full
finite source solution to the lens equation, \ref{eq-mult_lens},
is required to determine the precise limits on the microlensing
planet detection method set by the finite angular size of the
sources. 

\begin{figure}
\centering
\includegraphics[height=6.5cm]{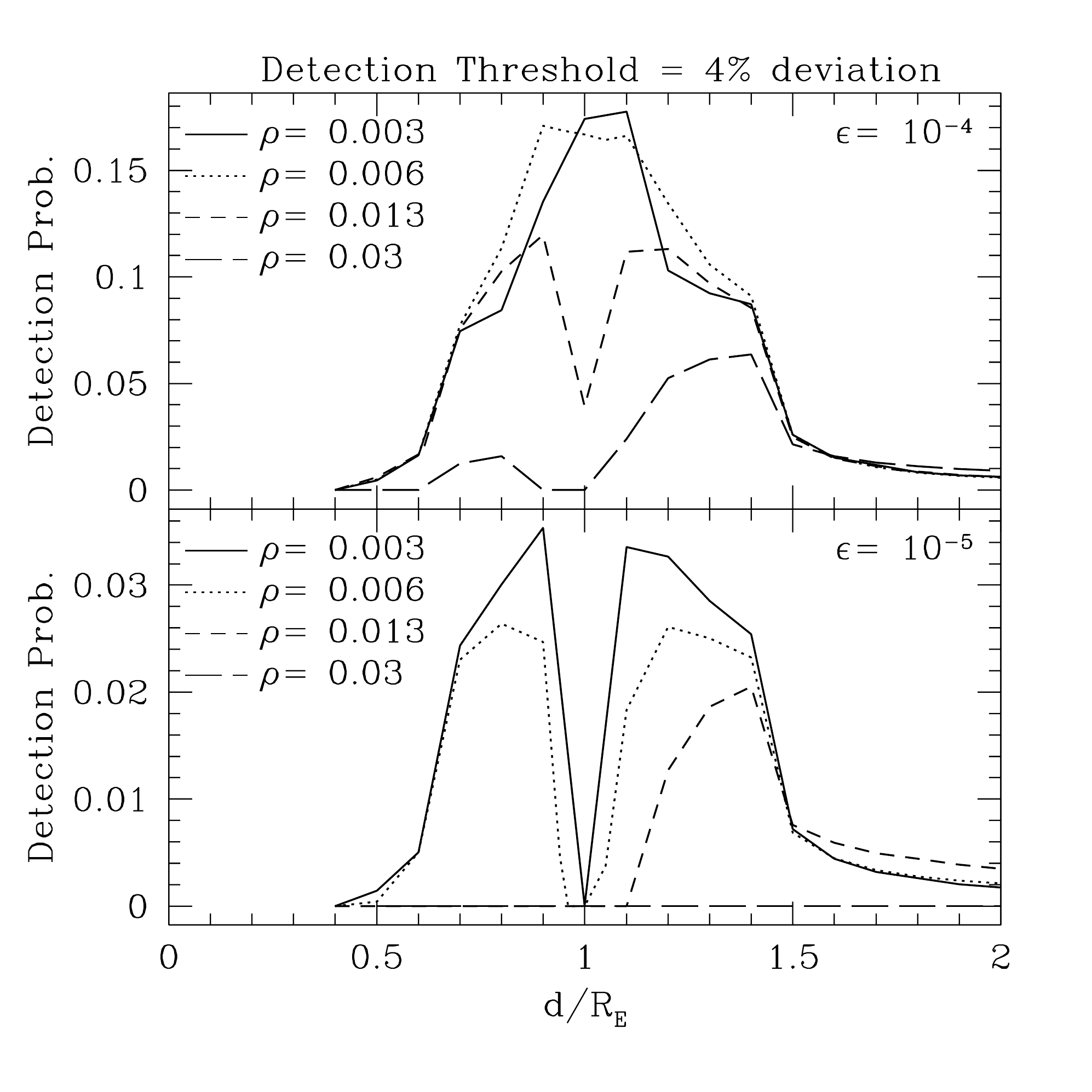}
%
%
\caption{The planetary deviation detection probability is plotted for different
  values of the planetary mass ratio, $\epsilon$, and the stellar radii,
  $\rho$ (in units of $R_E$. 
  A planet is considered to be ``detected" if the lightcurve deviates
  from the standard point lens lightcurve by more than 4\% for a duration of
  more than $t_E/200$.}
\label{fig-fin_src_prob}       
\end{figure}

Full finite source planetary microlensing light curves were
first calculated by \citet{em_planet} using the
methods described in \S~\ref{sec-multilens}. Results of these
calculations are reproduced in Figs.~\ref{fig-fin_src_lc} and
\ref{fig-fin_src_prob}. Fig.~\ref{fig-fin_src_lc} shows a series
of planetary light curves with planetary mass fractions of
$\epsilon = 10^{-4}$ and $10^{-5}$. For a typical lens star
mass of $\sim 0.3\msun$, these correspond to $1\mearth$ and
$10\mearth$, respectively. The finite source light curves
are characterized by the source star radius in
Einstein ring units: $\rho \equiv \theta_\ast/\theta_E$.
The $\rho$ values shown in Fig.~\ref{fig-fin_src_lc} are
0.003, 0.006, 0.013, and 0.03, and these span the expected
range of $\rho$ for a low mass planetary host star in the
Galactic bulge with a source star ranging in radius from
$1\rsun$ to $10\rsun$, which is a typical radius for a
``red clump" K-giant in the bulge. A number of the planet 
detections to date actually have $\rho$ values in the
0.4-1$\times 10^{-3}$ range because the lens stars
reside in the bulge and have a larger than average mass.
(Both of these imply a larger $\theta_E$.)

Several general trends are apparent from Fig.~\ref{fig-fin_src_lc}.
First, the planetary deviations are easily detectable for 
$\rho = 0.003$, but the signals are much weaker for $\rho = 0.03$.
This implies that $1\mearth$ planets are easily detected with
main sequence source stars, but planets of $10\mearth$
are close to the lower limit of detectability for giant source
stars. Another notable feature of these light curves is that
the planetary signals with $d = 0.8$ are more easily 
washed out by the finite source effects than those
with $d = 1.3$. This is a consequence of the large
magnification deficit seen between the two planetary
caustics for a minor image perturbation, as shown in the
left hand panel of Fig.~\ref{fig-ob71_caus}. When a large
finite source effectively averages over the vicinity of a
minor image planetary caustic, the positive and negative
deviations effectively cancel each other out
\citep{em_planet,gould-gauch}. In contrast,
the major image planetary caustic magnification deviation pattern
is mostly positive, so the finite source effect merely smoothes it
out.

Fig.~\ref{fig-fin_src_prob} shows how the planet detection
probability varies as a function of $d$ for the same $\epsilon$
and $\rho$ values used for Fig.~\ref{fig-fin_src_lc}.
The greater tolerance of deviations with $d > 1$ 
to finite source effects is apparent for $\epsilon = 10^{-4}$,
$\rho = 0.03$ and $\epsilon = 10^{-5}$, $\rho = 0.013$.
The behavior of Fig.~\ref{fig-fin_src_prob} near $d=1$
is even more interesting. For $\epsilon = 10^{-4}$ and
$\rho = 0.003$ or 0.006, the detection probability reaches a
maximum at $d \approx 1$, but for $\rho = 0.013$, the
probability has a local minimum at $d=1$, and for $\rho = 0.03$
or any of the $\rho$ values with $\epsilon = 10^{-5}$, the
detection probability $=0$ at $d=1$. This is due to the fact
that for $d\approx 1$ the planetary and stellar caustics merge for
form a relatively large single caustic that is extended along
the lens axis. This caustic is large, but relatively weak, and it
has associated positive and negative deviation regions that
tend to cancel when averaged over by a moderately
large finite source.

However, some features of Fig.~\ref{fig-fin_src_prob}
are dependent on the somewhat arbitrary choice of 
the event detection threshold, and the sensitivity of
a real observing strategy can differ from this. In fact,
the planetary deviation detected in event
OGLE-2005-BLG-169 \citep{ogle169} would not
have passed the selection criteria for Fig.~\ref{fig-fin_src_prob},
but the planet is nevertheless detected with a strong 
signal. The reason for this is that it was identified as
a very high magnification event with a very high sensitivity
to planets, and for this reason it was observed much
more frequently than most events with potential
planetary signals. The additional observations
provided enough additional signal to allow the
definitive detection of a relatively low-amplitude
signal.

As a practical matter, finite source effects imply a lower
planetary mass limit of $M_p \simgt 5\mearth$ for giant
source stars in the bulge, and a limit of $M_p \simgt 0.05\mearth$
for bulge main sequence stars. Thus, searches for 
terrestrial exoplanets must focus on main sequence source
stars.
 
\section{Planetary Parameters from Microlensing Events}
\label{sec-params}

The determination of the properties of the lens systems
that are detected in microlensing events is often a serious 
challenge. The simple form of the microlensing light curves
shown in Fig.~\ref{fig-lcex} is an advantage when trying to
identify microlensing events, but as
I mentioned in \S~\ref{sec-singlelens}, in a single lens
event, it can also be a drawback when trying to interpret
observed microlensing events. For most single lens events, it is
only the $t_E$ parameter that constraints the physically interesting 
parameters of the event: the lens mass, $M$, the lens distance, $D_L$,
and the relative velocity, $\vperp$. The single lens parameters
$u_0$ and $t_0$ don't constrain lens system parameters that
are of much interest.

In addition to the parameters needed to describe a single lens
event, a planetary microlensing event must have three additional
binary lens parameters: the planetary mass ratio,
$q = \epsilon/(1-\epsilon)$, the star-planet separation, $d$,
(which is in units of $R_E$), and the angle between
the star-planet axis and the trajectory of the source
with respect to the lens system, $\theta$. So, two of these
new parameters, $q$ and $d$, directly constrain 
planetary parameters of interest, although $d$ is
normalized to $R_E$, which may not be known. Most planetary
light curves, at least those for low-mass planets, also have
caustic crossings or a close approach to a cusp that reveal
light curve features due to the finite size of the source star.
This enables the source radius crossing time, $t_\ast$, to
be measured.

The determination of the star-planet separation and the
planetary mass fraction is usually quite straightforward 
from the microlensing light curve. For events at moderate
magnification, due to the planetary caustic, the separation
can be determined by the magnification predicted by the 
single lens model that describes the event outside the
region of the planetary deviation following eq.~\ref{eq-d_sc}.
This still leaves an ambiguity between the $d<1$ and $d>1$
solutions, but this is easily resolved by the drastically different
magnification patterns in the vicinity of major image and
minor image caustics, as shown in Fig.~\ref{fig-ob71_caus}.
The planetary mass fraction, $q$, can generally be 
determined by the duration of the planetary perturbation.
In some cases, if the time scale of the deviation is similar to 
or smaller than $t_\ast$, both $q$ and $t_\ast$ determine
the deviation time scale, but good light curve coverage with
moderately precise photometry allow both $q$ and $t_{\ast}$
to be determined \citep{gaudigould_plpar}.

The situation is somewhat different for high magnification, stellar 
caustic deviation events. \citet{dominik99} pointed out an
approximate degeneracy in the properties of the stellar
caustic under the transformation $d\rightarrow 1/d$, which 
means that there may be a $d\leftrightarrow 1/d$ ambiguity 
in the modeling of stellar caustic planetary events. This is
apparent from the magnification patterns shown in 
Fig.~\ref{fig-ob71_caus}. For as source trajectory nearly
parallel to the lens axis or for 
$d\sim 1$, this degeneracy breaks down, so the ambiguity
disappears. With precise photometry it is usually possible to
to distinguish between the $d<1$ and $d>1$ solutions, and
this has been the case for all events observed to date.

\subsection{Angular Einstein Radius}
\label{sec-theta_E}

A large fraction of planetary light curve deviations exhibit
finite source effects that allow the source radius crossing time,
$t_\ast$, to be measured. This is the case for most detectable
events with a planetary mass, $M_p \simlt 10\mearth$, but 
for gas giant planets of $M_p \simgt 300\mearth$, it is possible
to detect a planetary deviation without the source crossing
a caustic or coming close enough to a cusp to display finite
source effects. So, $t_\ast$ is measurable for most,
but not all planetary microlensing events.

When $t_\ast$ is measured, it is possible to place an
additional constraint, as long as the angular radius of
the source star, $\theta_\ast$, can be estimated, because
the angular Einstein radius is given by
\begin{equation}
\theta_E = {\theta_\ast t_E\over t_\ast } \ .
\label{eq-th_E_th_s}
\end{equation}

The angular radius of the source star can be measured if 
the brightness and color of the source are known with the
use of empirical color-angular radius relations 
\citep{vanbelle,kervella_dwarf}. In the crowded fields where
microlensing events are observed, the most reliable measure
of the source star brightness and color comes from the 
light curve models, which include the source brightness as
a model parameter. So, although it is sufficient to measure the
detailed light curve shape in a single passband, it is important
to obtain a few measurements during the microlensing event
in at least one additional passband so that the light curve fit
will also reveal the color of the source. It is also important to
estimate the extinction towards the source. With measurements
in only two colors, such as $V$ and $I$, the extinction
can be estimated by comparison to the red clump
giant stars within an arc minute or two of the target
star \citep{yoo_rad}. While this does not yield a precise measure
of the extinction to the source, note that an error in the extinction
to the source will affect both the estimated intrinsic brightness
and color of the source. Fortunately, the extinction-induced 
brightness and color errors have the opposite effect on the 
estimated source star radius. This partial cancelation implies
that the estimated $\theta_\ast$ value is not very sensitive to
to the uncertainty in the extinction.

A more precise estimate of $\theta_\ast$ can be obtained with
observations during the microlensing event in more than two
passbands, particularly if one of the passbands is in the infrared
because the optical-IR color-radius relations are much more precise
than the optical ones \citep{kervella_dwarf}
and because extinction is much lower in
the IR than in the optical. Observations in 3 or more colors also
allow an estimate of extinction that doesn't depend the
nearby clump giants, with the use of empirical color-color
relations \citep{bessellbrett}.

\begin{figure}
\centering
\includegraphics[height=8cm]{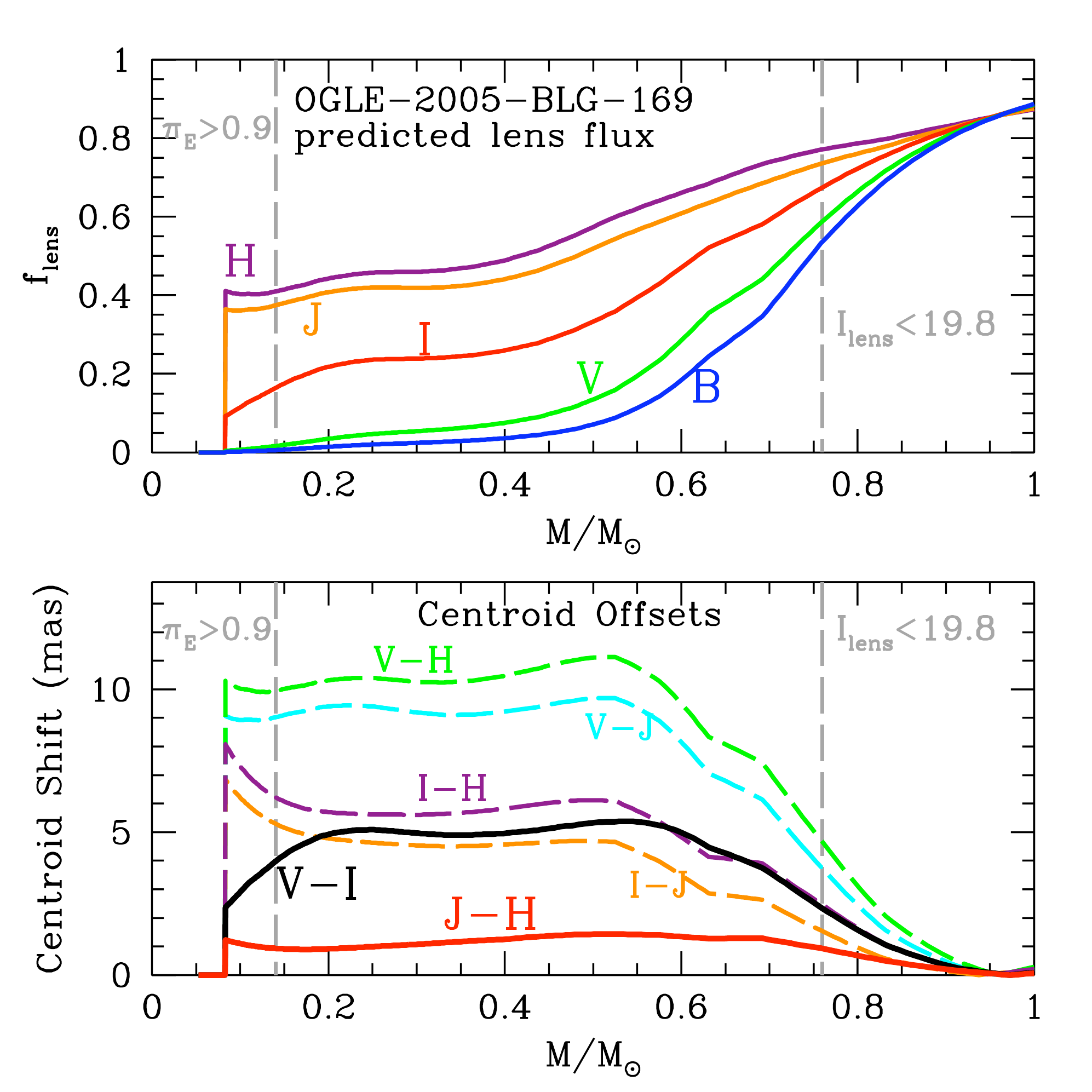}
%
%
\caption{The predicted fractional brightness, $f_{\rm lens} = F_L/( F_S + F_L)$,
of the OGLE-2003-BLG-169 lens is plotted in the top panel
as a function of mass in the
$BVIJH$ passbands. ($F_L$ and $F_S$ and the detected flux of the 
lens and source, respectively. The predicted offsets of the centroids
of the blended source$+$lens images in different passbands are shown
in the bottom panel, assuming that the images are taken 2.4 years after
peak magnification.}
\label{fig-ogle169_hst}       
\end{figure}

When the angular Einstein radius is measured, we have the following relation,
\begin{equation}
M_L = {c^2\over 4G} \theta_E^2 {D_S D_L\over D_S - D_L} \ ,
\label{eq-m_thetaE}
\end{equation}
which can be considered to be a mass-distance relation because $D_S$ is
generally known (approximately) from the brightness and color of
the source. (The high density of stars in the Galactic bulge means that the
source is almost always a bulge star.)
Eq.~\ref{eq-m_thetaE} provides a one-parameter family of solutions to the 
microlensing event, and this can be converted to a measurement of the 
planetary host star properties with one additional piece of information.
Since the brightness of the source star can be determined by the
light curve fit, the brightness of the lens star can be determined with an
image that has sufficient angular resolution to resolve the source and 
lens stars from the unrelated stars in the field. This generally requires
space-based imaging with the Hubble Space Telescope (HST), or
possibly ground-based adaptive optics imaging because of the 
extreme crowding in the Galactic bulge fields where microlensing events
are most easily found. (The lens-source relative proper motion has 
typical value $\mu_{\rm rel}\sim 5\,$mas/yr, so the lens and source are
not typically resolved from each other until a decade or more after the
event.) If the combined lens-plus-source image is significantly brighter
than the brightness of the source from the microlensing fit, then the 
difference determines the brightness of the lens. This then allows
the mass of the planetary host (lens) star to be determined using
a main sequence star mass-luminosity relation \citep{plan_char}.

The top panel of Fig.~\ref{fig-ogle169_hst} shows the predicted
brightness of the lens for the OGLE-2005-BLG-169 event in the
$BVIJH$ passbands. This indicates that the lens star will easily
be detected if it is a main sequence star, since even a
$0.08\msun$ lens star will contribute
$\simgt 40$\% of the $H$-band flux and $\simgt 10$\% of the
$I$-band flux. This case is more favorable than most because of
a relatively large $\theta_E$ value, but most cases, the lens star
will be detectable in the $H$-band unless it is a late M-dwarf 
located in the bulge. However, for Galactic disk lenses at 
a certain range of distances
(corresponding to $0.2\msun \simlt M \simlt 0.4$ for 
OGLE-2005-BLG-169 in the $IJH$-bands) the mass-distance relation, 
eq.~\ref{eq-m_thetaE}, combines with the mass-luminosity
relation to yield a nearly flat mass-brightness relation for the
planetary host star. In these cases, it is useful to have
images in shorter wavelength bands, such as $V$ and $B$
because this cancelation does generally not occur in the 
optical and infrared passbands for the same range of 
lens star masses.

\begin{figure}
\centering
\includegraphics[height=6cm]{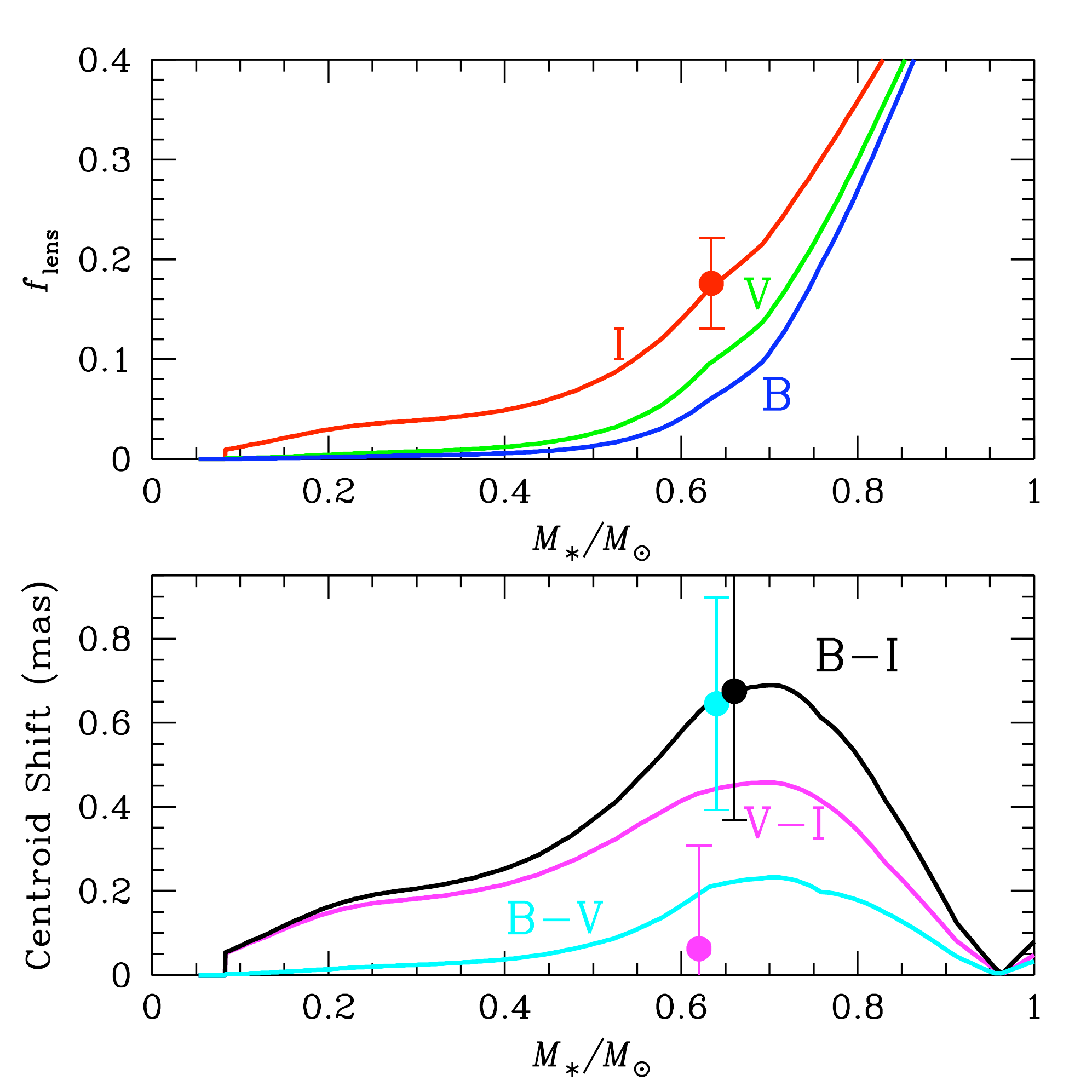}
\includegraphics[height=6cm]{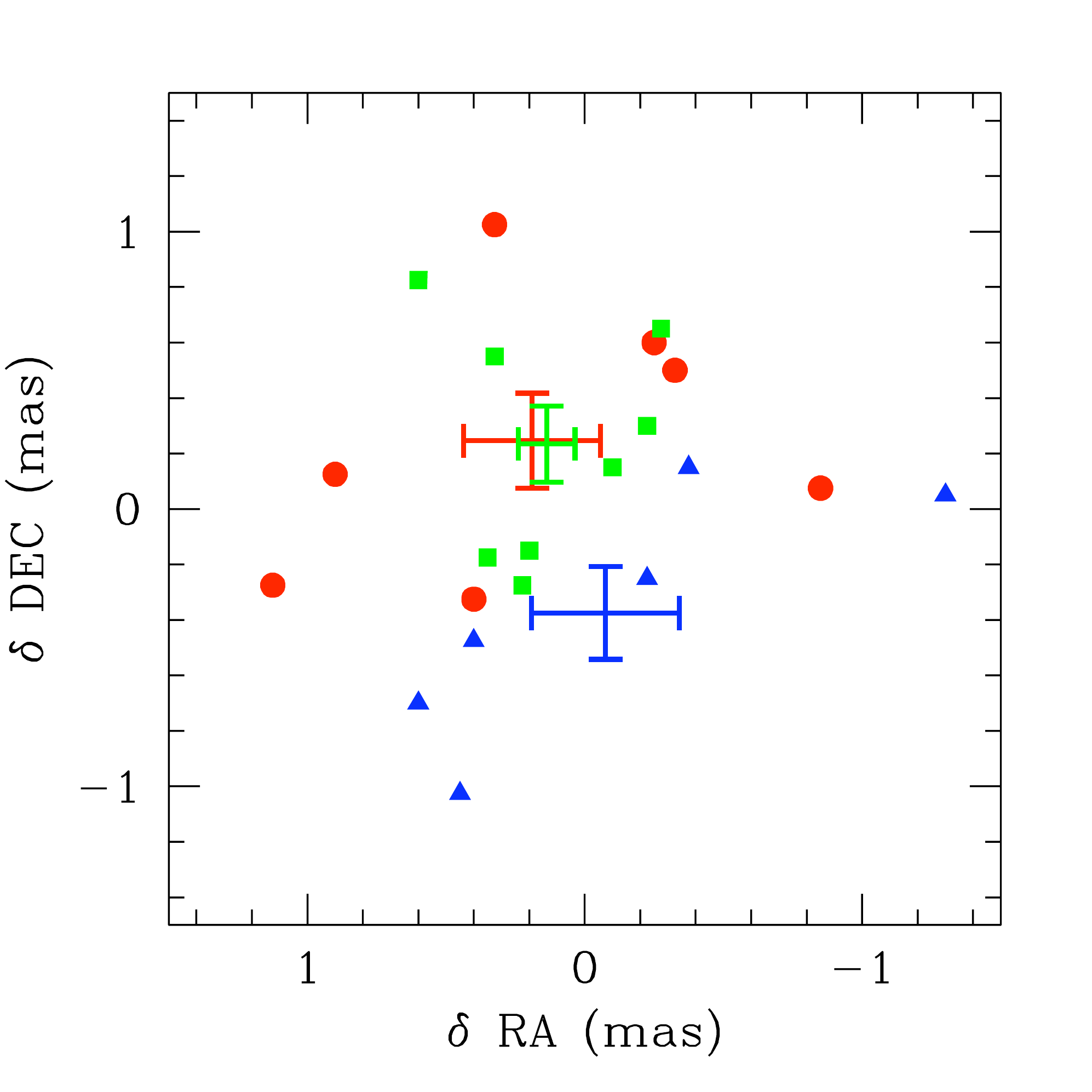}
\caption{The top-left panel shows the fraction of the source$+$lens flux 
for event OGLE-2003-BLG-235/MOA-2003-BLG-53 that is predicted to
come from the lens in the HST-$I$, $V$, and $B$
passbands as a function of lens mass. The bottom-left panel
shows the predicted color-dependent centroid shifts as a function
of mass for 1.78 years of relative proper motion at $\mu_{\rm rel} = 3.3\,$mas/yr.
The measured values of $f_{\rm lens}$ in the I-band and the color dependent
centroid shifts and error bars are indicated with their error bars. These
are plotted at an arbitrary value for the stellar mass ($M_\ast$).
The centroids of the source$+$lens star blended images
in the individual HST/ACS/HRC images are shown in the right panel as 
red circles ($I$), green squares ($V$), and
blue triangles ($B$). The crossed error bars are the
average centroid in each passband.
}
\label{fig-moa53_hst}       
\end{figure}

High resolution images in multiple colors also allow an independent
method for estimating the lens star brightness, as shown in the bottom
panel of Fig.~\ref{fig-ogle169_hst} and the bottom-left panel of
Fig.~\ref{fig-moa53_hst}. Because the lens and source stars
usually have different colors, the centroid of the blended 
source$+$lens image will usually be color dependent. So, an
additional constraint on the lens star is obtained by measuring
the centroid offset between the centroids of the blended source$+$lens
in different passbands. As indicated in Fig.~\ref{fig-moa53_hst}, this
effect was marginally detected for the first planet detected by 
microlensing \citep{bennett-moa53} with HST images taken only
1.8 years after peak magnification. Also, because this color
dependent centroid shift depends on the relative lens-source
proper motion, $\mu_{\rm rel}$, it can be used to help determine
$\theta_E$ for planetary events with no finite source effects,
and hence, no measurement of $t_\ast$.

\begin{figure}
\centering
\includegraphics[height=8.5cm]{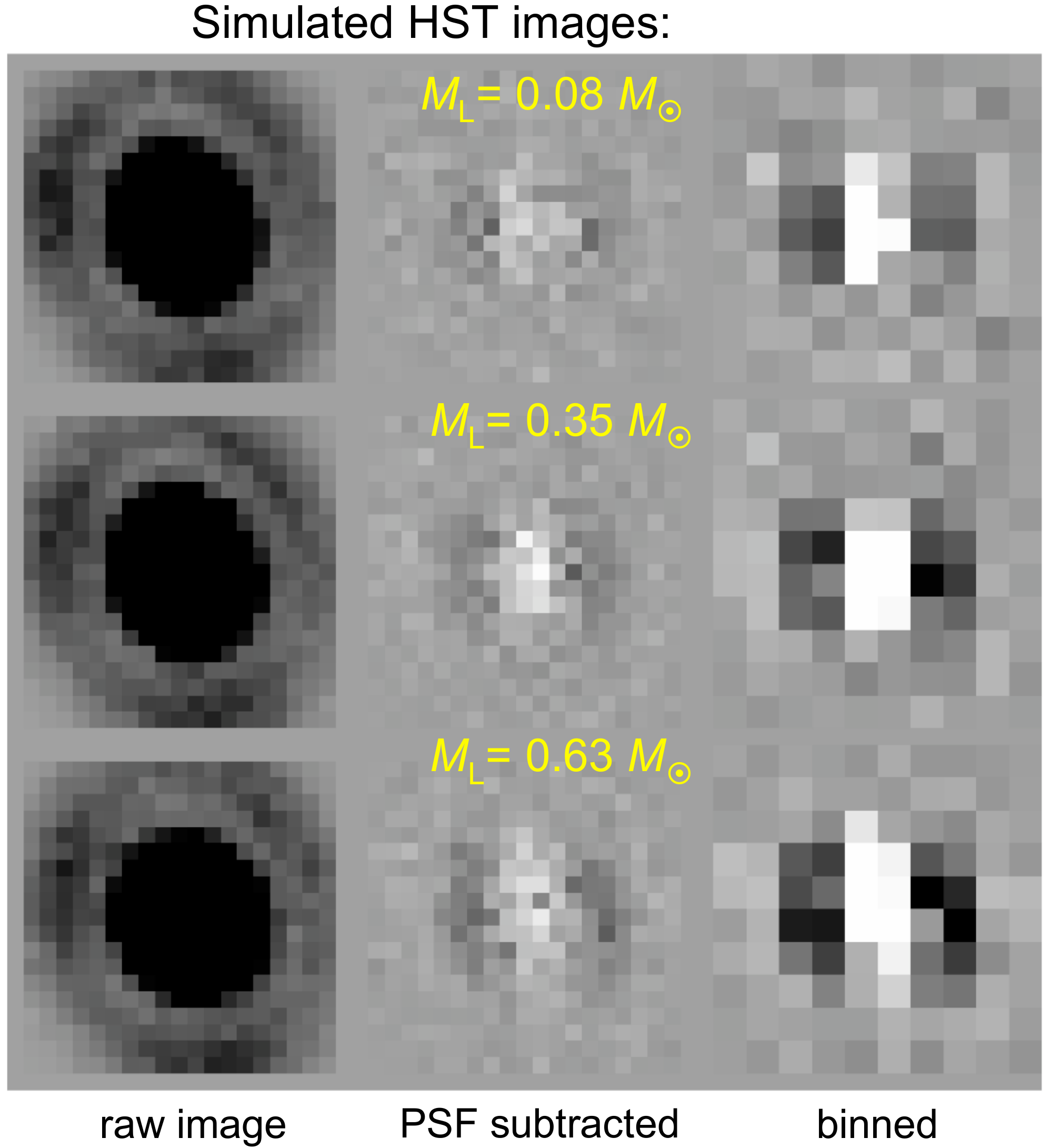}
\caption{
Simulated image stacks of multiple dithered exposures of the 
OGLE-2005-BLG-169 source and lens star 2.4 years after peak magnification
using the HST/ACS High Resolution Camera (HRC) in the F814W filter 
band. The top row of images assumed a host star mass of 
$M_\ast = 0.08\msun$, the middle row assumes $M_\ast = 0.35\msun$, and 
the bottom row assumes $M_\ast = 0.63\msun$. In each row, the image 
on the left shows the raw image stack sampled at one half the native 
HRC ($28\,$mas) pixel size. The central column shows the residuals after subtraction 
of the best fit PSF model, showing the blended image elongation along 
the $x$-axis due to the lens-source 
separation. The right hand column shows these residuals
rebinned to the $28\,$mas native pixel scale.
\label{fig-pics}}
\end{figure}

The stable point-spread function (PSF) of space-based telescopes,
such as HST, allows the measurement of the image elongation
due to the growing separation of the lens and source stars
after the microlensing event. Simulations of this effect for
the OGLE-2005-BLG-169 event are shown in Fig.~\ref{fig-pics}
for three different cases: $M_L = 0.08\msun$, $M_L = 0.35\msun$, 
and $M_L = 0.63\msun$. This event has a higher relative proper motion
than most events, but this simulation assume images taken only
2.4 years after peak magnification, so for other events, it may be
necessary to obtain the follow-up space-based images $\sim 4$ years
after peak magnification.

When $\mu_{\rm rel}$ can be measured from image elongation
and/or the color dependent centroid shift, then the angular Einstein
radius can be determined via
\begin{equation}
\theta_E = \mu_{\rm rel} t_E \ ,
\label{eq-thetaE_mu}
\end{equation}
so that mass-distance relation can be determined even when
$t_\ast$ cannot be measured.

\subsection{Microlensing Parallax}
\label{sec-par}

Another method to ``solve" a microlensing event involves the
microlensing parallax effect. This 
refers to measurements of $\rep = R_E D_S/(D_S-D_L)$, the
Einstein radius projected to the position of the Solar System.
$\rep$ can be measured with the help of 
observations of microlensing events by observers
at different locations. Because the Einstein radius is typically
of order $R_E\sim 1\,$AU, the observers must generally
be separated by a distance of $\sim 1\,$AU. The conceptually
simplest way to do this is to observe an event simultaneously 
with a satellite in a heliocentric orbit, \citep{refsdal-par,gould-par1},
as has recently been done with Spitzer \citep{dong-ogle05smc1}.
However, it is much more common to use the orbital motion
of the Earth the measure the microlensing parallax effect
\citep{macho-par1,mao-par,ogle-par2,bennett-parbh,ogle-99b22},
but for events of very high magnification, it is possible to measure
this effect with observations from different observatories on
Earth \citep{gould_ext_ML97}, as has recently been done by
Gould \etal (2007, in preparation).

A potential complication with this method is that the orbital motion
of the source star can mimic the effect of the orbital motion of the
Earth, but if the signal is strong, it is generally possible to 
detect the characteristic features of the Earth's orbit
\citep{multi-par}. Another potential complication is that for
events with $t_E \ll 1\,$yr, it is often possible to measure only
a single component of the two-dimensional ${\bf \rep}$ vector
\citep{smith-par_acc}.
But, for events with detectable lens stars, the two-dimensional
relative proper motion, ${\mathbold \mu_{\rm rel}}$, it is
possible to determine the full ${\bf \rep}$  vector because
${\bf \rep} \parallel  {\mathbold \mu_{\rm rel}}$.

When $\rep$ and $\theta_E$ are both measured, the lens
system mass is given by
\begin{equation}
M_L = {c^2\over 4G} \rep \theta_E \ .
\label{eq-m_rep}
\end{equation}
This method has been used to determine the lens mass for
a binary star lens system towards the Galactic bulge \citep{planet-er2000b5}
and a low-mass M-dwarf lens towards the Large Magellanic Cloud
\citep{lmc5-mass}. The first use of this method in a planetary
microlensing event is the case of the double planet event
OGLE-2006-BLG-109, to be published later this year
(Gaudi \etal\ 2007, in preparation; Bennett \etal\ 2007, in
preparation).

\subsection{Planetary Orbits}
\label{sec-orb}

The final property of a planetary system that can be measured is the orbital
motion of the planet with respect to the star. This is a lower order effect
than microlensing parallax because we see the effects of both the 
planet and the star in the light curve. So, we are sensitive to the relative
velocity between the star and planet, whereas the velocity of the Earth
around the Sun cannot be separated from the lens-source relative
velocity. However, the time scale of the planetary deviation is generally
only a small fraction of the microlensing light curve, and this
limits the amount of time over which we can detect the orbital
motion effects. Also, the typical orbital period of a planet detected
by microlensing is $\sim 10\,$yrs, so the orbital velocities are 
generally lower than that of the Earth around the Sun.

The for a planetary deviation of duration $\Delta t$ and an orbital
period, $P$, the orbital motion during the planetary deviation
causes a shift in the planetary lens position with respect to the
source of order
\begin{equation}
\Delta u \approx \Delta t {2\pi \over P} \approx  0.002-0.02 \ ,
\label{eq-dt_orb}
\end{equation}
assuming a planetary deviation duration of $1-10\,$days.
In oder to determine whether eq.~\ref{eq-dt_orb} indicates
that the effect of orbital motion is detectable, we need to know
what value of $\Delta u$ is measurable. On thing that limits
our resolution in $\Delta u$ is the finite angular size of the 
source star. The typical angular size for a bulge main sequence
source is $\theta_\ast \sim 0.5\,\mu$as and a typical angular Einstein radius
for a bulge event is $\theta_E \sim 0.5\,$mas, so the source radius is
typically of order $\rho = \theta_\ast /\theta_E \sim 0.001$. So, if
we can detect $\Delta u$ as small as $0.1\rho$, then we could
be sensitive to $\Delta u \sim 10^{-4}$. 

In practice, it can be difficult to do this well in the measurement of
orbital effects because changes in other model parameters can
often compensate for the change in $\Delta u$ due to orbital
motion. In order to retain a constraint on the orbital motion, it
is generally necessary to have a relatively complicated planetary
deviation with more than a single caustic crossing or cusp passage
that is well sampled by the data. Finally, for events with relatively
long planetary signals, the orbital acceleration can be as large as
$\Delta u \approx (\Delta t 2\pi/P)^2 \approx 4\times 10^{-4}$.
So, with a very well sampled planetary deviation it is also possible
to measure the orbital acceleration, as well as the velocity.

\section{Observational Programs}
\label{sec-obs_prog}

There are a variety of different observing programs that contribute to
the detection of planets via gravitational microlensing. The most basic
requirement is to be able to identify microlensing events, as was first
done by the MACHO Collaboration towards the LMC \citep{macho-nat93} 
and OGLE group toward the Galactic bulge \citep{ogle93}.
Because microlensing
observing programs do not yet have the resources to observe $\simgt 10\,$square
degrees of the Galactic bulge several times per hour, it has been necessary
to follow a strategy first suggested by \citet{gouldloeb}. Stellar microlensing
events must be identified in progress, and then followed with a global network
of telescopes on an $\sim$hourly time scale. The MACHO 
\citep{macho-iauc1,macho-alert} and OGLE \citep{ogle-ews} groups developed
real-time microlensing detection systems within a year after the first
microlensing events were discovered, and this led to the first spectroscopic
confirmation of a microlensing event \citep{benetti-spec}. The MOA group
began real time detections in 2000 \citep{moa-alert} and was the first
group to employ real time event detection with the more advanced 
difference imaging photometry method \citep{moa-2002}.

The first microlensing follow-up projects were the 
Probing Lensing Anomalies NETwork or PLANET group
\citep{planet_95season} and the Global Microlensing Alert
Network, or GMAN, \citep{macho-gman95}, which both began taking
data in 1995. The PLANET team followed the \citet{gouldloeb} strategy,
but the GMAN group focused more on non-planetary microlensing.
A second follow-up group focused on exoplanets, the Microlensing
Planet Search (MPS) collaboration began in 1997 \citep{mps-98smc1},
but MPS merged with PLANET in 2004. The final microlensing follow-up
group is the Microlensing Follow-up Network or MicroFUN 
\citep{yoo_rad}, which began observations in 2003. MicroFUN
does not follow the  \citet{gouldloeb} strategy, but instead focuses on
high magnification microlensing events as suggested by
\citet{griest_saf}.

\subsection{Early Observational Results}
\label{sec-early_obs}

The most definitive of the early planetary microlensing observational
results involved limits on the presence of planets based on the 
lack of detection of planetary signals. The MPS and MOA groups
reported the first planetary limits from a high magnification
event \citep{mps-98blg35}. This, was the first demonstration
of sensitivity to Earth-mass planets by any method, except for
pulsar timing \citep{pulsar_planets}.
The PLANET group followed with limits
from a lower magnification event \citep{planet-o98bul14} and then
a systematic analysis of five years worth of null detections
\citep{planet-limit,gaudi-planet-lim}. They found that less than 33\% of 
the lens stars in the inner Galactic disk and bulge have companions 
of a Jupiter mass or greater between 1.5 and 4 AU. These papers 
claim that their limits apply to Galactic bulge M-dwarfs, but this
summary of the PLANET result
neglects an important bias in the events that have
been searched for planets. The microlensing teams are more
efficient at finding long time scale microlensing events
\citep{macho-bulge-diff,moa-tau,macho-clump-tau,ogle2-blg-tau,eros-blg-tau}
The long events are also more likely to be discovered prior to peak
magnification, so they can be more efficiently searched for planetary
signals. As a result, the median time scale of the events search for
planets in \citet{gaudi-planet-lim} is $\VEV{t_E} = 37\,$days, while the
actual efficiency corrected median time scale is  $\VEV{t_E} = 16\,$days.
This implies that the events that have been searched for planets have more
massive lens stars and are more likely to reside in the disk than the
typical Galactic bulge microlensing event. Thus, it is probably the case
that most of the events searched by \citet{gaudi-planet-lim} have lens
stars that are more massive than an M-dwarf, reside in the Galactic
disk, or both.

In addition to these upper limits on the planetary frequency, there
were also a number of less-than-certain planet detections. 
\citet{rhie_lmc1} showed that the very first microlensing event
discovered showed a light curve feature that could be explained
by a planet, but there
was a near equal mass binary lens fit that could also explain the
data. The MACHO group
\citep{macho_planets} pointed out that there is a good chance
that event MACHO-95-BLG-3 was caused by a free-floating 
Jupiter-mass planet.
The MPS group found that their data for MACHO-97-BLG-41
was best explained by a Jupiter-mass planet orbiting a
binary star system \citep{mps-97blg41}, but the PLANET data
for this event favored an orbiting binary star interpretation
\citep{planet-97blg41}. (Some of the MPS data are now known to
be contaminated by moonlight reflecting off the telescope optics.)
An analysis by the MOA group \citep{moa-himag-obs} showed that
the combined MACHO, MOA, MPS, and PLANET data for
MACHO-98-BLG-35 was consistent with the low S/N detection
of a terrestrial planet. Finally, \citet{jar_pac_ob02055} showed that
the event OGLE-2002-BLG-55 had a signal consistent with
a planet detection, but \citet{gaudi_ob02055} pointed out that
there were other possible explanations.

\begin{table}
\centering
\caption{Exoplanets Discovered by Microlensing}
\label{tab-planets}
\begin{tabular}{lcccl}
\hline
\noalign{\smallskip}
Event Name & \ Star Mass \ & \ Planet Mass\  & \ Semi-Major Axis \  & Lead Group  \\
\noalign{\smallskip}
\hline
\noalign{\smallskip}
OGLE-2003-BLG-235Lb/ & $0.63 {+0.07\atop -0.09}\msun$ & 
     $830 {+250\atop -190} \mearth$ & $4.3 {+2.5\atop -0.8}$AU & MOA \\
\ \ \ MOA-2003-BLG-53Lb & & & \\
 \vspace{0.1cm}
OGLE-2005-BLG-71Lb & $0.46\pm 0.04 \msun$ & 
     $1100 \pm 100  \mearth$ & $4.4\pm 1.8\,$AU & OGLE \\
 \vspace{0.1cm}
OGLE-2005-BLG-390Lb & $0.22 {+0.21\atop -0.11}\msun$ & 
     $5.5{+5.5\atop -2.7} \mearth$ & $2.6 {+1.5\atop -0.6}$AU & PLANET \\
 \vspace{0.1cm}
OGLE-2005-BLG-169Lb & $0.49{+0.14\atop -0.18}\msun$ &
     $13{+4\atop -5} \mearth$ & $3.2{+1.5\atop -1.0}$AU & MicroFUN \\
OGLE-2006-BLG-109Lb & $0.50\pm 0.05 \msun$ & 
     $226\pm 25 \mearth$ & $2.3\pm 0.2$AU & MicroFUN \\
OGLE-2006-BLG-109Lc & $0.50\pm 0.05 \msun$ &
     $86\pm 10 \mearth$ & $4.6\pm 0.5$AU & MicroFUN \\
\noalign{\smallskip}
\hline.
\end{tabular}
\end{table}

\subsection{Microlensing Planet Detections}
\label{sec-obs_detections}

\begin{figure}
\centering
\includegraphics[height=10cm]{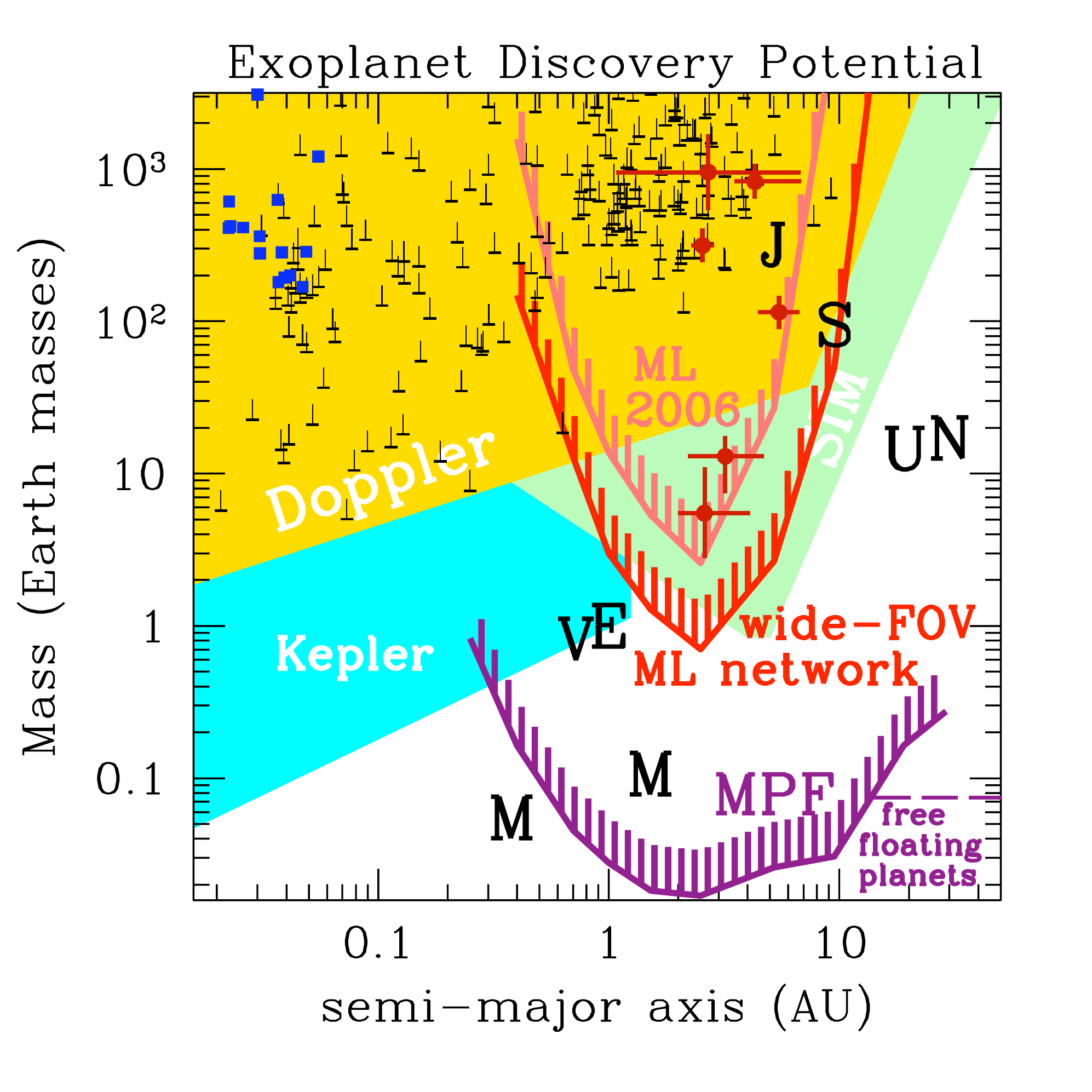}
%
%
\caption{The sensitivity of various exoplanet detection methods is
plotted in the mass vs. semi-major axis plane. Doppler radial velocity 
detections are shown in black, with 1-sided error bars for the $m \sin i$
uncertainty. Planets first detected by transits are shown in blue, and
the microlensing planet discoveries are shown in red. The gold, cyan
and green shaded
regions show the sensitivity of the radial velocity method and NASA's
Kepler and SIM missions, respectively. The light red and red curves
show the sensitivity of current and future microlensing planet search
programs, and the purple curve gives the sensitivity of the proposed
Microlensing Planet Finder (MPF) mission.}
\label{fig-m_v_sep}       
\end{figure}

Table~\ref{tab-planets} summarizes the properties of the planets discovered by
microlensing to date, including
four published microlensing exoplanet discoveries 
\citep{bond-moa53,ogle71,ogle390,ogle169} plus a 2-planet system that will 
soon be published (Gaudi \etal\ 2008, Bennett \etal\ 2009, in  preparation).
The microlensing discoveries 
are compared to other known exoplanets in Fig.~\ref{fig-m_v_sep}.

\begin{figure}
\centering
\includegraphics[height=6cm]{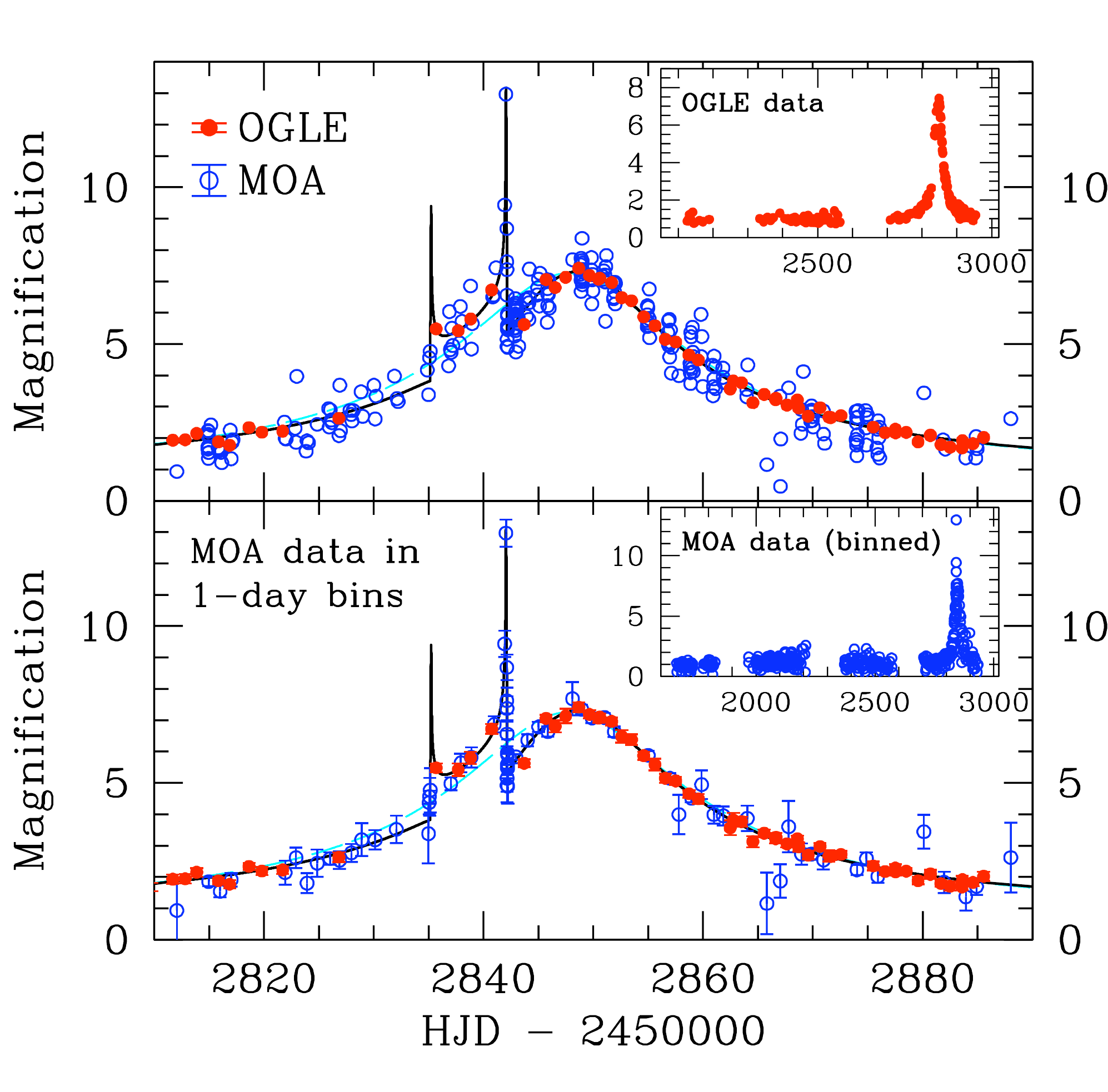}
\includegraphics[height=6cm]{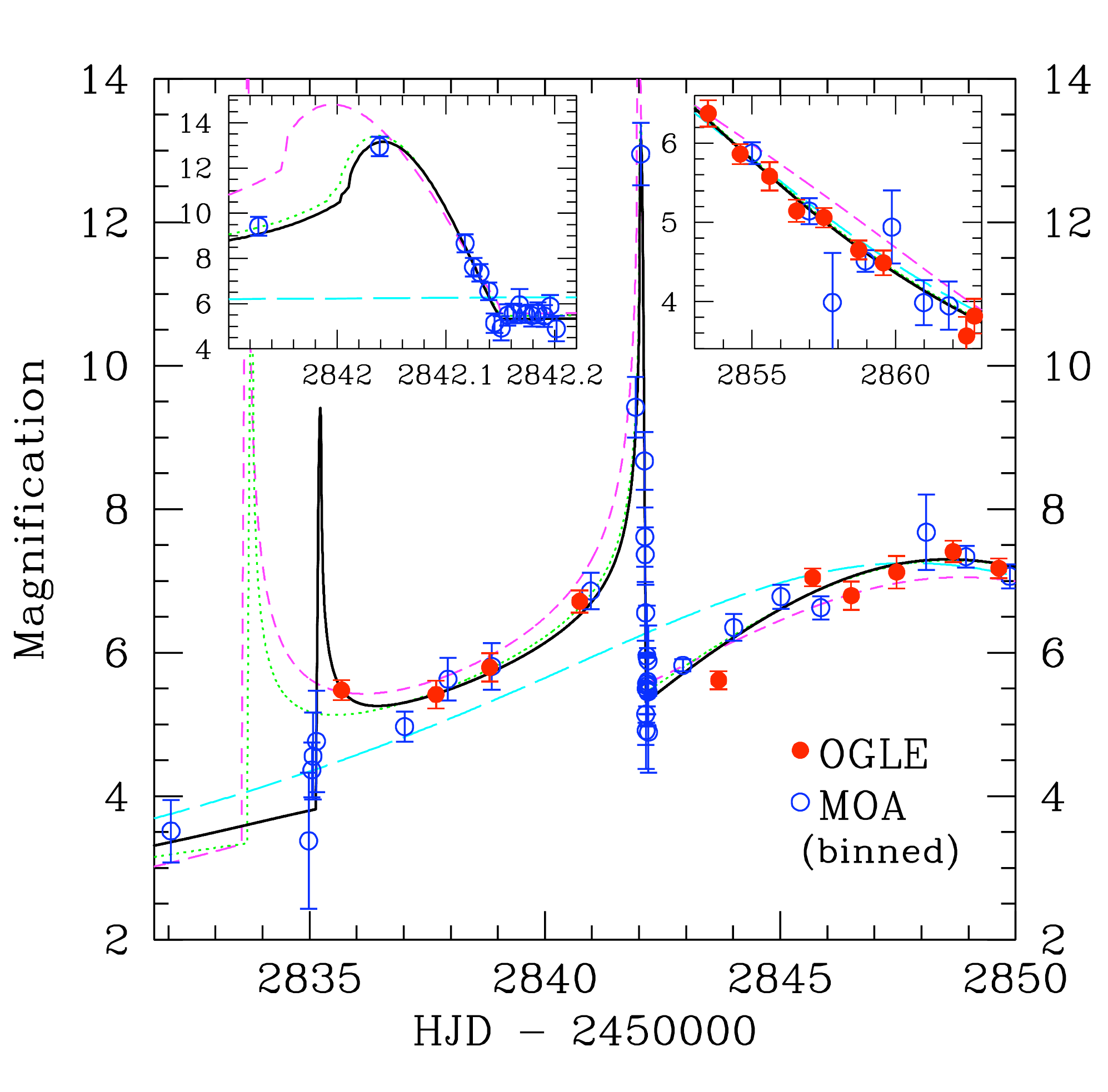}
%
%
\caption{The OGLE-2003-BLG-235/MOA-2003-BLG-53 light curve with
OGLE data in red and MOA data in blue.
The top-left panel presents the complete data set during
2003 (main panel) and the 2001Ð2003 OGLE data (inset). 
The median errors in the OGLE and MOA points
are indicated in the legend. The bottom panel is the same as the top panel,
but with the MOA data grouped in 1 day bins, except for the caustic crossing
nights, and with the inset showing MOA photometry during 2000Ð2003. The
binary- and single-lens fits are indicated by the
solid black and cyan dashed curves, respectively. The right panel shows the
light curve and models during caustic traverse.
These models are the single-lens
case ({\it cyan, long-dashed curve}), the best binary lens with $q \simgt 0.03$ 
({\it magenta, short-dashed line}), the planetary lens with caustic entry before day 2835 
({\it green, dotted line}), and the best overall fit with $q = 0.0039$({\it black, solid line}). 
The insets show the second caustic crossing and a region of the declining part of
the light curve where the best-fit nonplanetary binary-lens model fails to fit
the data. MOA data on days other than the caustic entry and exit (days
$2835\pm 0.5$ and $2842 \pm 0.5$) are placed in 1 day bins.
}
\label{fig-moa53_lc}       
\end{figure}

The first planet discovered by microlensing is shown in Fig.~\ref{fig-moa53_lc}.
The light curve is plotted in units of the source star flux, which is 
determined by the best microlensing model to the event, because the
star field is too crowded to determine the unmagnified stellar flux directly.
This event was first discovered by the OGLE group and announced via
their ``early warning system" as event OGLE-2003-BLG-235
on 2003 June 22. On 2003 July 21, the alert system of the MOA-I
microlensing survey detected this event and reported it as
MOA-2003-BLG-53. The MOA detection came later because the
MOA-I telescope had only a $0.61\,$m aperture and has worse
seeing conditions than are typical at the $1.3\,$m OGLE telescope
in Chile. However, the MOA telescope had a larger field-of-view (FOV),
and this enabled them to image each of their survey fields $\sim 5$
times per clear night. As a result, MOA was able to detect the
second caustic crossing for this event, and arrange for the
additional observations that caught the caustic crossing endpoint
(thanks to first author, Ian Bond, who was monitoring the photometry
in real time).

The naming convention for planets discovered is that the name from the
first team to find the microlensing event is used for the event, so in this
case OGLE-2003-BLG-235 takes precedence over MOA-2003-BLG-53.
When referring to the lens system, we add a suffix ``L", and when referring
to the source, we add an ``S". For a lens or source system that is multiple,
we add an additional capital letter suffix for a stellar mass object or a lower
case letter for a planetary mass companion. So, OGLE-2006-BLG-109LA,
 OGLE-2006-BLG-109Lb, and  OGLE-2006-BLG-109Lc, refer to the star and
 two known planets of the OGLE-2006-BLG-109 lens system. This convention
 provides names for multiple components of the source star system. For example,
 OGLE-2022-BLG-876Sb would refer to a planetary companion to the 
 source star (which would be difficult, but not impossible 
\citep{graff-gaudi,ML-hot-jup}to detect.

It is interesting to note that this event was discovered by a procedure that differs
from both the alert-plus-followup strategy suggested by \citet{gouldloeb} 
and the high magnification strategy suggested by \citet{griest_saf}.
Instead, the planetary deviation was detected in the observations of
one of the survey teams, and identified in time to obtain additional data
to confirm the planetary nature of the light curve deviation. We will return
to this strategy later in the discussion of future microlensing projects
given in \S~\ref{sec-future}.

\begin{figure}
\centering
\includegraphics[height=8cm]{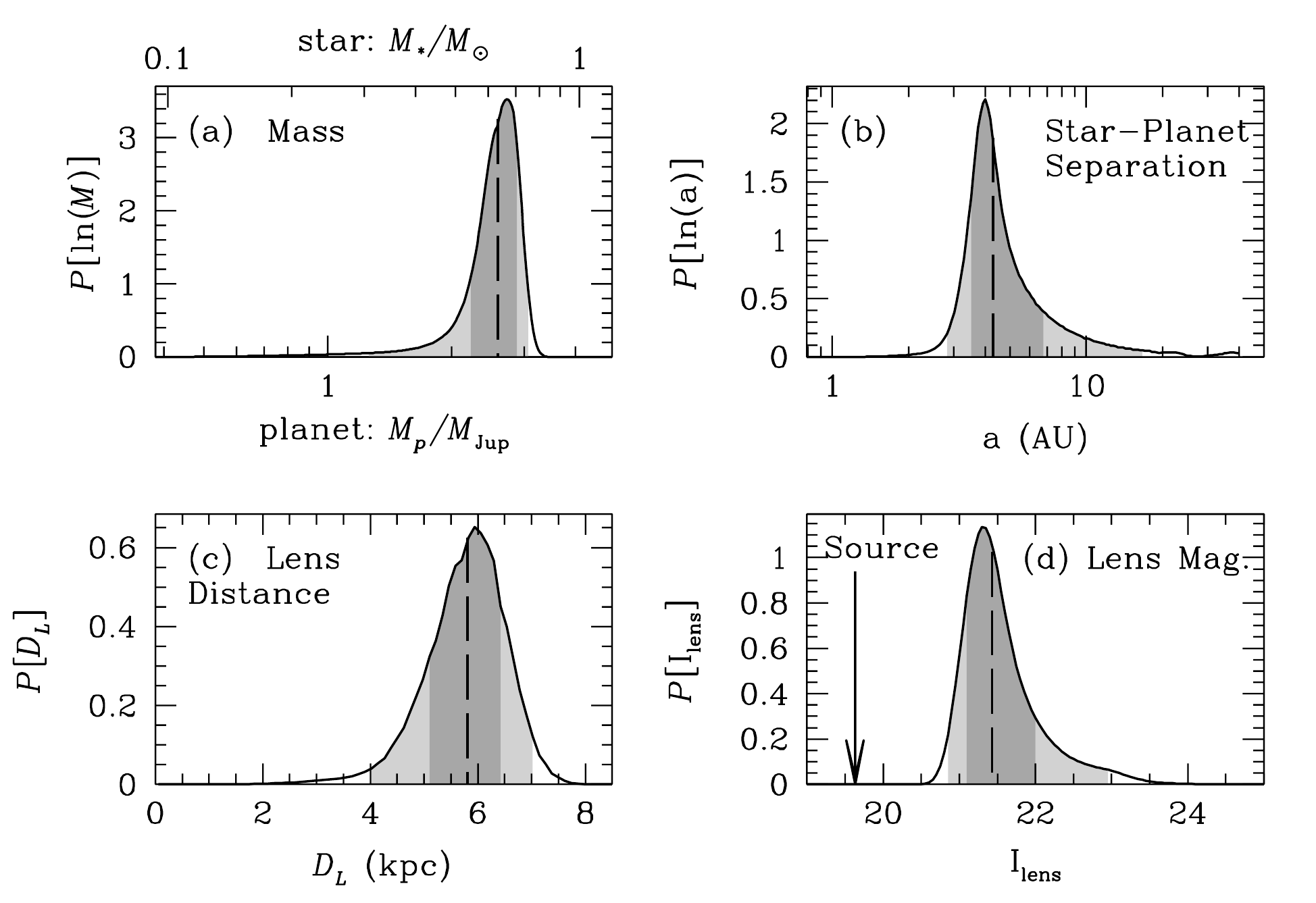}
%
%
\caption{Bayesian probability densities for the properties of the planet,
OGLE-2003-BLG-235Lb,
and its host star if it is a main sequence star. 
(a) The masses of the lens star and its planet ($M_\ast$
and $M_p$ respectively). 
(b) the separation, 
(c) their distance from the observer ($D_L$); 
and (d) the I-band brightness of the host star. 
The dashed vertical lines indicate the medians, and the shading
indicates the central 68.3\% and 95.4\% confidence intervals.  All
estimates follow from a Bayesian analysis assuming a standard model for the
disk and bulge population of the Milky Way, the stellar mass function of
Bennett \& Rhie (2002).}
\label{fig-moa53_prop}       
\end{figure}

Another notable feature of this event is that the lens star has been
identified in HST images \citep{bennett-moa53}. 
As indicated in Fig.~\ref{fig-moa53_hst},
there is an extra source of light superimposed at the location of the
source star. This is very likely to be the lens star, and if so, the HST
photometry implies that a fraction, $f_{\rm lens} = 0.18\pm 0.05$, of the
total source plus lens flux comes from the lens. During the
microlensing event, the lens and source were separated by $< 0.1\,$mas,
but by the time of the HST images, $\Delta t =1.78\,$years after peak magnification,
the lens-source separation should have grown to 
$\Delta t \mu_{\rm rel} = 5.9\pm 0.7\,$mas. 
($\mu_{\rm rel} = 3.3\pm 0.4\,$mas/yr was determined from eq.~\ref{eq-th_E_th_s}
with input parameters from the light curve model.) This separation, plus the
mass-distance relation, eq.~\ref{eq-m_thetaE}, enable to derivation of the
curves shown in the bottom left panel of Fig.~\ref{fig-moa53_hst}. These
show the amplitude for the offset  of the centroids of the blended 
lens plus source images in different color bands. The HST data
indicate a marginal detection of this color-dependent centroid shift
at a level consistent with the assumption that the excess flux is
due to the lens. 

With this marginal detection of the color-dependent centroid shift,
we can't be absolutely sure that the lens star has been detected
because it is possible that the excess flux could be due to a companion
to the source star. It is straight forward to deal with this uncertainty
with a Bayesian analysis \citep{bennett-moa53}, and the results of
such an analysis are shown in FIg.~\ref{fig-moa53_prop}. The
resulting most likely parameter values for the event parameters
are a host star mass of $M_\ast = 0.63{+0.07\atop -0.09}\msun$,
a planet mass of $M_{\rm p} =  2.6 {+0.8\atop -0.6} M_{\rm Jup}$,
and an orbital semi-major axis of $a = 4.3 {+2.5\atop -0.8}\,$AU.
The distance to the lens system is $D_L = 5.8 {+0.6\atop -0.7}\,$kpc, and
the lens star magnitude is $I_L = 21.4 {+0.6\atop -0.3}$.

\begin{figure}
\centering
\includegraphics[height=12cm]{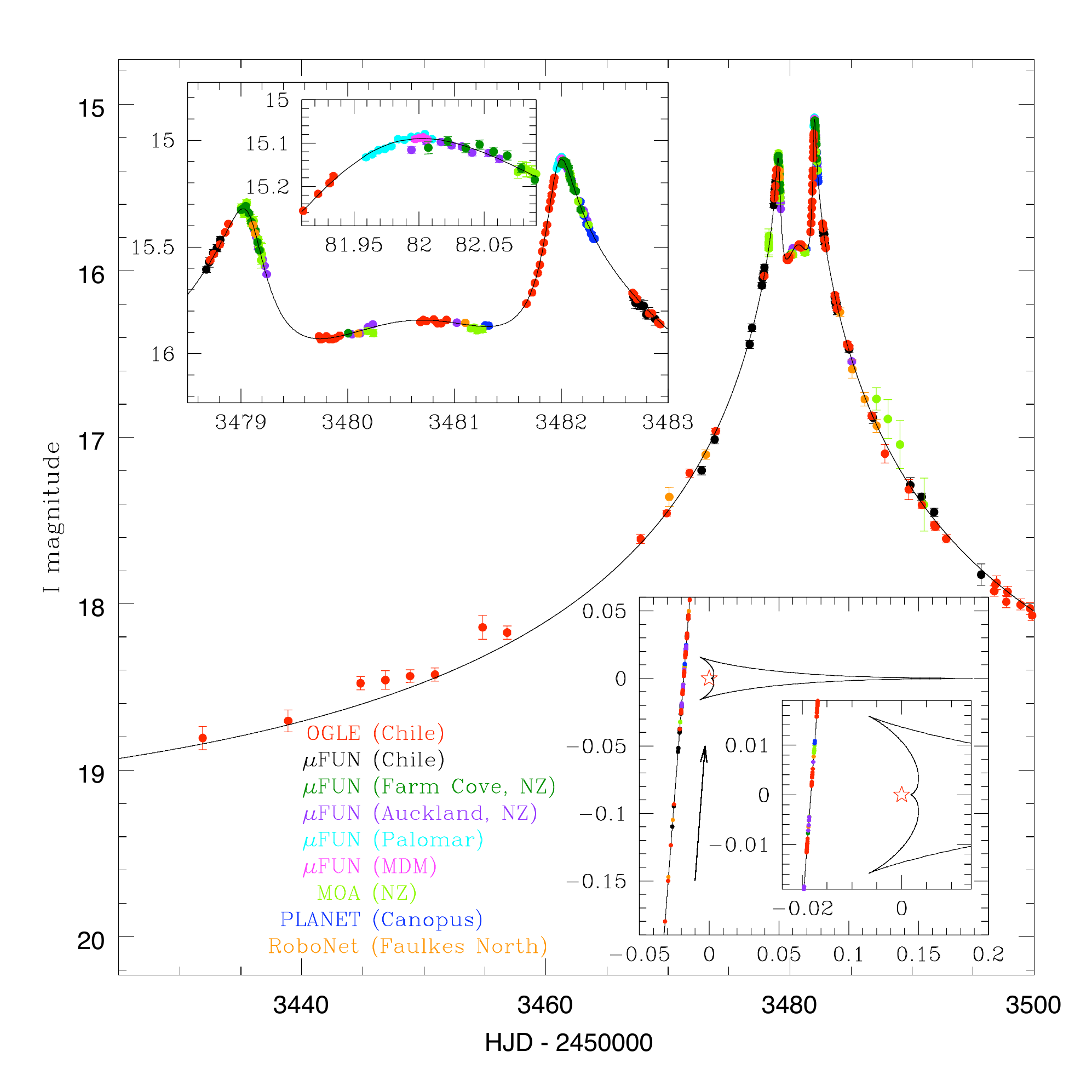}
%
%
\caption{The OGLE-2005-BLG-71 light curve showing the planetary anomaly 
near thepeak.The triple peak (two large symmetric peaks surrounding a small peak)
indicates that the source passed three cusps of a caustic, the middle one
being weak ({\it insets}). The interval between peaks (and so cusps) 
is $\Delta t = 3\,$days, implying that the companion mass must be small.}
\label{fig-ogle71_lc}       
\end{figure}

The light curve of the second planet discovered by microlensing,
OGLE-2005-BLG-71Lb, is shown in Fig.~\ref{fig-ogle71_lc}
\citep{ogle71}. This
was a moderately high magnification event that would have reached
a maximum magnification of $A_{\rm max} \approx 42$ if the lens star
had no planets. Because of the $d\leftrightarrow 1/d$ ambiguity
discussed in \S~\ref{sec-params}, this event has two models
that explain the major features of the light curve quite well.
Fig.~\ref{fig-ob71_caus} shows the magnification patterns for
these models, and for the trajectory of the lens, which is nearly
perpendicular to the lens axis, the light curves for these 
different models are quite similar. From \citet{ogle71}
the physically interesting parameters of the best fit models are
$t_E = 70.9\pm 3.3$, $q = 7.1\pm 0.3 \times 10^{-3}$, and
$d = 1.294 \pm 0.002$ for the ``wide" model and
$t_E = 73.9\pm 3.5$, $q = 6.7\pm 0.3 \times 10^{-3}$, and
$d = 0.758 \pm 0.002$ for the ``close" model. However, the
$\chi^2$ difference between these two models is
$\Delta \chi^2 = \chi^2_{\rm close} - \chi^2_{\rm wide} = 22.0$, 
so the ``wide" model is strongly preferred.

OGLE-2005-BLG-71Lb was the first planet discovery
with significant contributions from amateur astronomers,
with critical observations near the two strong cusp
approach peaks by Grant Christie of the Auckland
Observatory and Jennie McCormick of the Farm Cove
Observatory.

\begin{figure}
\centering
\includegraphics[height=8cm]{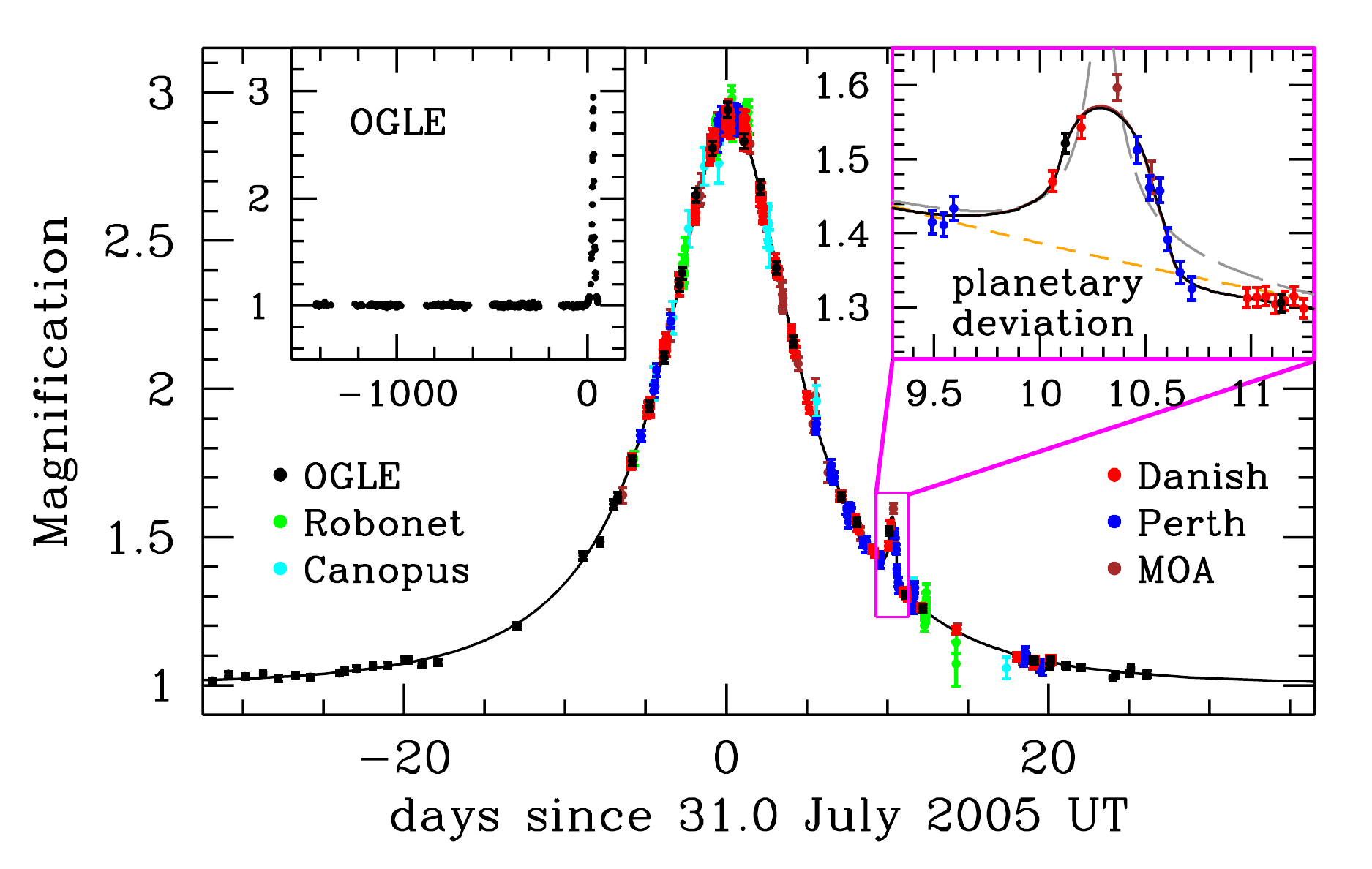}
%
%
\caption{
The observed light curve of the OGLE-2005-BLG-390
microlensing event and best-fit model plotted as a function of time. 
The data set consists of 650 data points from PLANET Danish
({\it red points}), PLANET Perth ({\it blue}), PLANET Canopus
(Hobart, {\it cyan}), RoboNet Faulkes North ({\it green}), OGLE 
({\it black}), and MOA ({\it brown}). The top left inset
shows the OGLE light curve extending over the previous 4 years, whereas the
top right one shows a zoom of the planetary deviation, covering a time
interval of 1.5 days. The solid curve is the best binary lens model described in
the text with $q = 7.6\pm 0.7 \times 10^{-5}$,
and a projected separation of $d = 1.610\pm 0.008 R_E$.
The dashed grey curve is the best binary source
model that is rejected by the data, and the dashed orange line is the best
single lens model.}
\label{fig-ogle390_lc}       
\end{figure}

With a mass ratio of $q = 7.1\pm 0.3 \times 10^{-3}$, OGLE-2005-BLG-71Lb
must certainly be a gas giant planet, but without further information
such as measurement of finite source effects,
the detection of the lens star or a measurement of the
microlensing parallax effect, we cannot determine the properties
of the host star or the planetary mass with much precision. 
Fortunately, we are able to detect lens star in a set of 
HST images, and the light curve yields weak detections
of both a finite source size and the microlensing parallax
effect. So, we expect to determine the host star
and planet masses and to convert their separation into
physical units, but this analysis is not yet complete
(Dong et al 2007, in preparation).

The first low-mass planet discovered by microlensing was
OGLE-2005-BLG-390Lb \citep{ogle390}, led by the PLANET Collaboration.
This planet is currently tied with Gl 581c \citep{udry-gl581cd} as the lowest mass
exoplanet orbiting a normal star yet to be discovered\footnote{
The minimum mass of $M_p \geq 5.03\mearth$ is often quoted for Gl 581c,
but the $M_p \sin i$ ambiguity of the radial velocity method implies that
the median predicted mass is $M_p = 5.5\mearth$. This is the appropriate
number to compare to other detection methods.}. This event was
detected through a planetary caustic deviation, and the amplitude of the
deviation was significantly reduced by the finite angular size of the
clump giant source star. If the planet were smaller by a factor of $\sim 2$,
it would not have been detected in this event. As originally pointed out
in \citet{em_planet} and discussed in \S~\ref{sec-finite_src}, a 
microlensing search for Earth-mass planets should focus on 
events with main sequence source stars.

\begin{figure}
\centering
\includegraphics[height=5.1cm]{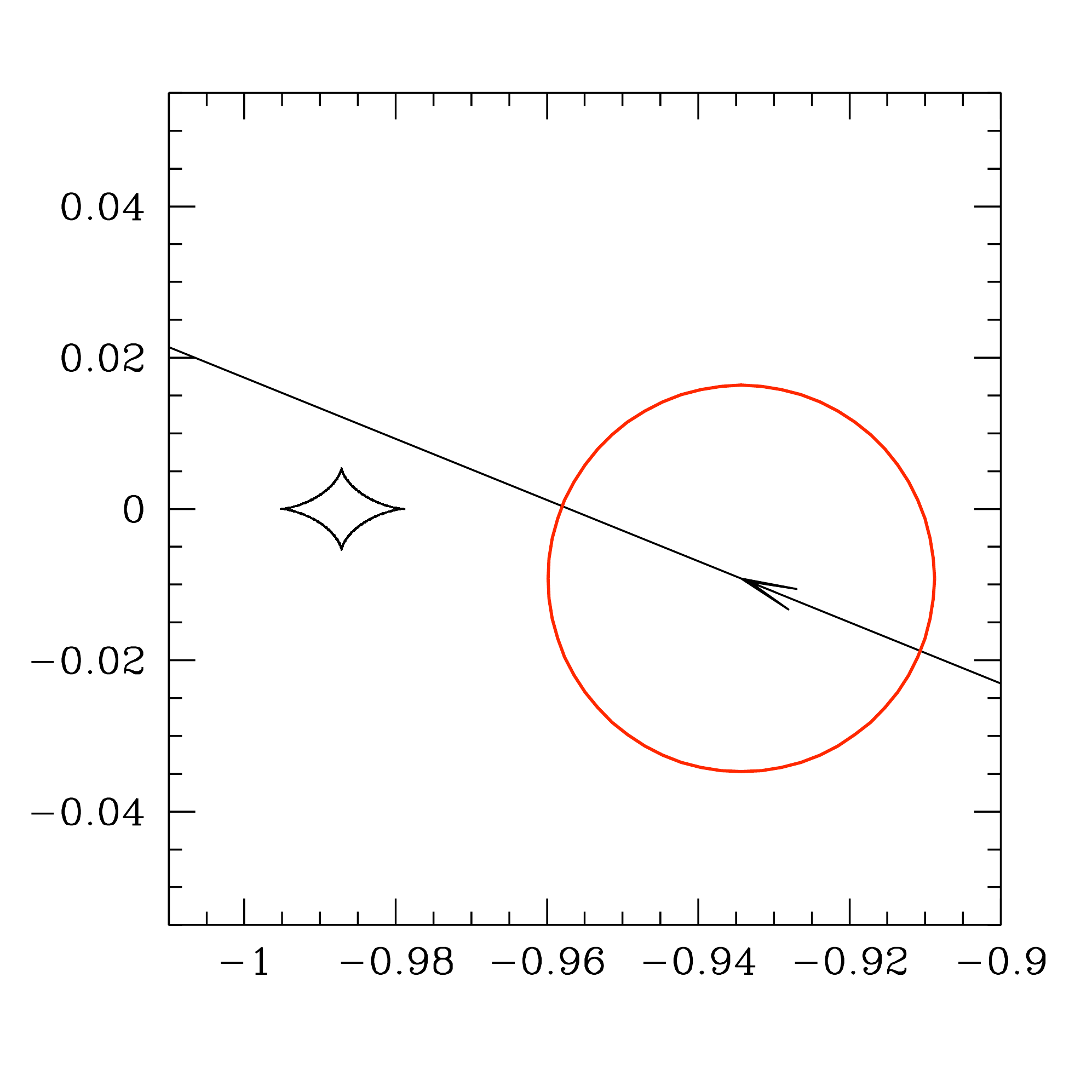}
\caption{
Comparison of the OGLE-2005-BLG-390Lb planetary caustic
({\it the black diamond shaped curve}) with the source star size
({\it red circle}). The black line with the arrow show the motion
of the source star.}
\label{fig-ogle390_caus}       
\end{figure}

The OGLE-2005-BLG-390Lb light curve
deviation does not show the characteristic features of a fold
caustic crossing, like OGLE-2003-BLG-235, or of a cusp approach,
like OGLE-2005-BLG-71. This is because the planetary caustic
is smaller than the source star's angular radius of 
$\theta_\ast = 5.3 \pm 0.7\mu$as is larger than the planetary 
caustic, as shown in Fig.~\ref{fig-ogle390_caus}. Because the
light curve does not show these characteristic binary-microlensing
features, we must consider a non-planetary explanation for the
light curve involving the lensing of a binary source star
by a single star lens. \citet{gaudi_pl_binsrc}. However, as the 
Fig.~\ref{fig-ogle390_lc} shows, a binary source model is a
poor fit to the data, as it fails to account for the Perth and
Danish data near the end of the perturbation. Formally, the
binary source model increases the fit $\chi^2$ by 
$\Delta\chi^2 = 46.25$ with one fewer degree of freedom.
These data 
are also sufficient to avoid a possible degeneracy in the
planetary parameters for such events that occurs when
the wings of the deviation are poorly sampled
\citep{gaudigould_plpar}.

\begin{figure}
\centering
\includegraphics[height=6.5cm]{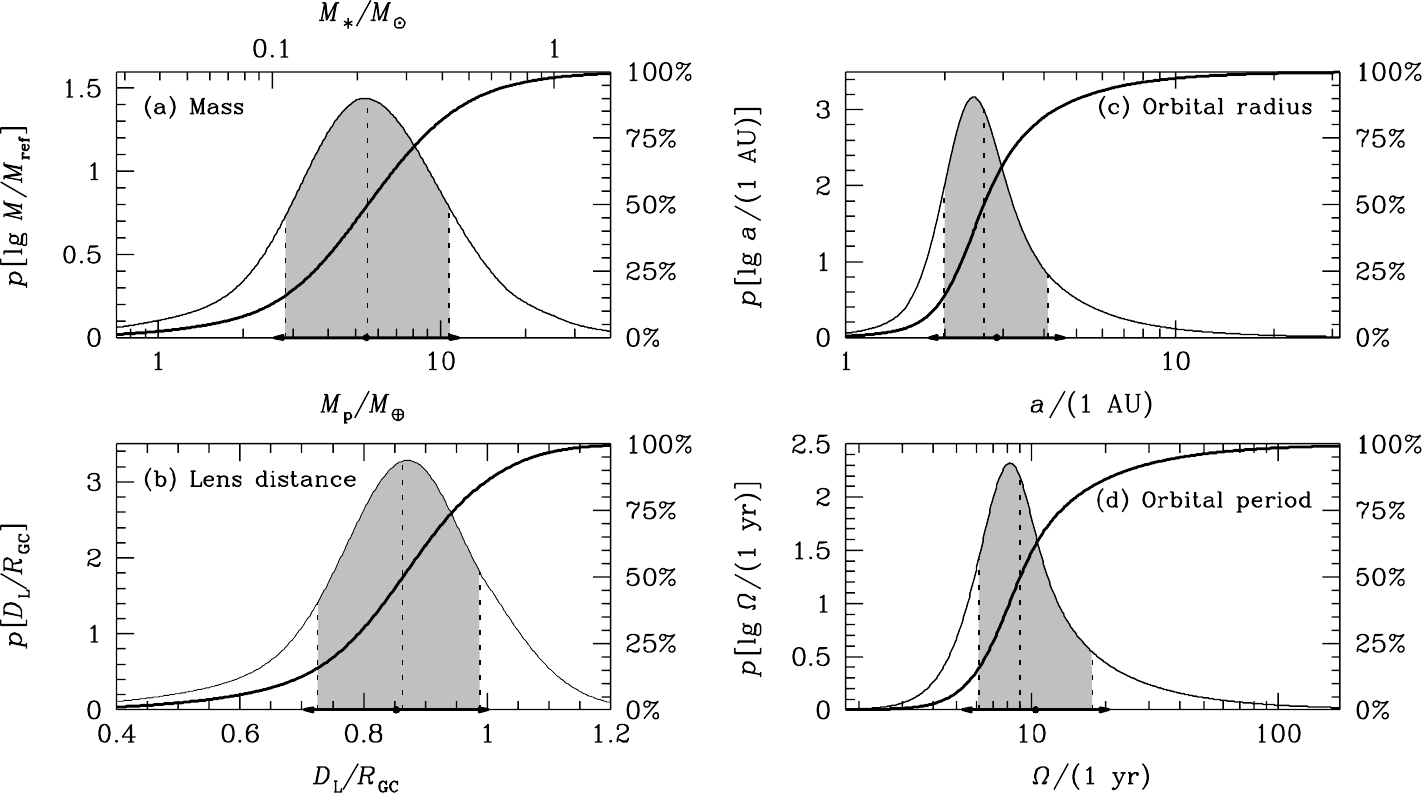}
%
%
\caption{
Bayesian probability densities for the properties of the planet
and its host star: (a), the masses of the lens star and its planet ($M_\ast$ and $M_p$
respectively), (b), their distance from the observer $D_L$, (c), the three
dimensional
separation or semi-major axis $a$ of an assumed circular
planetary orbit; and (d), the orbital period $\Omega$ of the planet.  
The bold, curved line in
each panel is the cumulative distribution, with the percentiles listed on the
right. The dashed vertical lines indicate the medians, and the shading
indicates the central 68.3\% confidence intervals, while dots and arrows on
the abscissa mark the expectation value and standard deviation. 
The medians of these distributions yield a $M_p = 5.5{+5.5\atop -2.7}\mearth$
planetary companion at a separation of $d = 2.6{+1.5\atop -0.6}\,$AU
from a $M_\ast = 0.22{+0.21\atop -0.11}\msun$
Galactic Bulge M-dwarf at a distance of $D_L = 6.6\pm 1.0\,$kpc
from the Sun. The median planetary period is $\Omega = 9{+9\atop -3}$years.}
\label{fig-ogle390_prop}       
\end{figure}

The microlensing model for this event directly determines
the planetÐstar mass ratio, $q = 7.6 \pm 0.7 \times 10^{-5}$, the projected
planetÐstar separation, $d = 1.610 \pm 0.008$, the Einstein radius
crossing time, $t_E = 11.03 \pm 0.11\,$days, and the source
radius crossing time, $t_\ast = 0.282 \times 0.010\,$days.
With the value for $\theta_\ast$ mentioned above, this yields the
angular Einstein radius, $\theta_E = 0.21\pm 0.03\,$mas, 
from eq.~\ref{eq-th_E_th_s} and the
mass-distance relation from eq.~\ref{eq-m_thetaE}. This mass-distance
relation can be combined with a standard Galactic model in a
Bayesian analysis to estimate the probability distribution 
of the lens system parameters 
\citep{macho-par1,macho-alert,multi-par,dominik06}. The results of
such an analysis are shown in Fig.~\ref{fig-ogle390_prop} following
the method of \citet{dominik06}, and nearly identical results are
obtained using the Galactic model and mass function parameters of
\citet{gest-sim}. This analysis gives a 95\%
probability that the planetary host star is a main-sequence star, a 4\%
probability that it is a white dwarf, and a probability of,1\% that it is
a neutron star or black hole.  The median parameters 
shown in Fig.~\ref{fig-ogle390_prop} imply that the planet
receives radiation from its host star that is only 0.1\% of the radiation
that the Earth receives from the Sun, so the probable surface
temperature of the planet is 50 K, similar to the temperature of
Neptune.

\begin{figure}
\centering
\includegraphics[height=11cm]{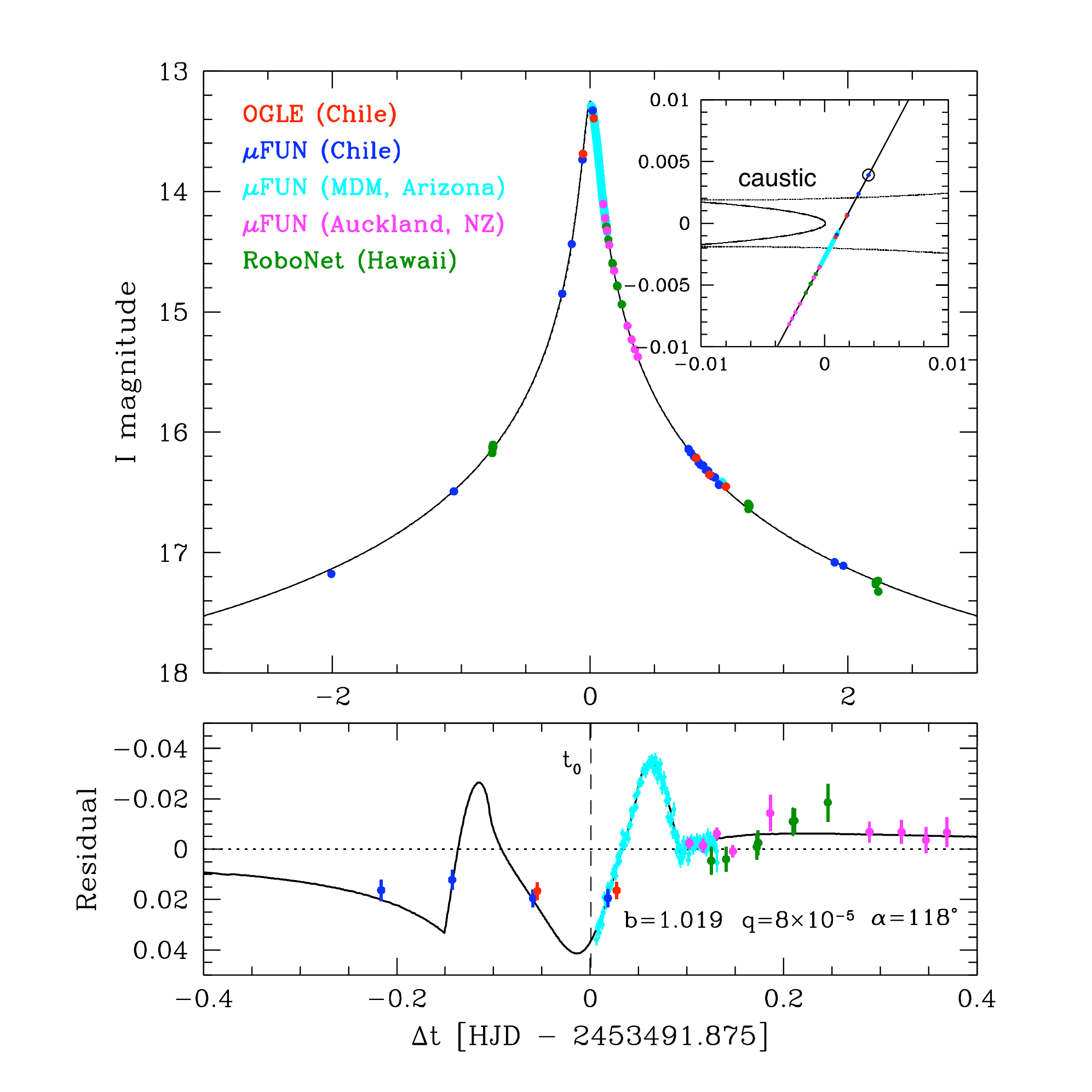}
%
%
\caption{
{\it Top}: Data and best-fit model for OGLE-2005-BLG-169. {\it Bottom}:
Difference between this model and a single-lens model with the same 
single lens parameters ($t_0$, $u_0$, $t_E$, and $\rho$).
It displays the classical form of a caustic entrance/exit that is often seen in
binary microlensing events, where the amplitudes and timescales are several
orders of magnitude larger than seen here. MDM data trace the characteristic
slope change at the caustic exit ($\Delta t = 0.092$) very well, while the entrance
is tracked by a single point at $\Delta t = -0.1427$. The dashed line indicates the time
$t_0$. {\it Inset}: 
Source path through the caustic geometry. The source size, $\rho$, is indicated.}
\label{fig-ogle169_lc}       
\end{figure}

As discussed in \S~\ref{sec-theta_E}, the lens star mass can be determined
directly if the lens star is detected. However, this will be quite
difficult for OGLE-2005-BLG-390L, because the source is a giant
star. For the median mass and distance to the lens system, the
lens star would be fainter than the source by a factor of 2000 in the
$K$-band. So, the detection of the lens star may require many years
for the relative proper motion of $\mu_{\rm rel} = 6.8\,$mas/yr and the 
development of new instruments for large ground-based or space
telescope.

OGLE-2005-BLG-169 was the third published event from the 2005 season
and the second low-mass planet found by microlensing
\citep{ogle169}. This was a very
high magnification event, with a peak magnification of $A_{\rm max} \simeq 800$,
and it's light curve is shown in Fig.~\ref{fig-ogle169_lc}. 
The bottom panel of Fig.~\ref{fig-ogle169_lc} indicates that the planetary
deviation has a maximum amplitude of about 4\% compared to the
light curve of the same event without a planet. Such low amplitude
deviations characteristic of the very weak caustics due to low-mass
planets near the Einstein ring. However, it is only part of the
caustic curve that is so weak. If the source would have passed
on the other side of the host star and crossed the backwards
``C" shaped part of the caustic in the inset of Fig.~\ref{fig-ogle169_lc},
the planetary signal would have been very much stronger. 
But, it order to detect the low amplitude signal due to the caustics
actually crossed by the source star, it was quite helpful to 
have continuous observations over the course of three hours
from the $2.4\,$m MDM telescope.

\begin{figure}
\centering
\includegraphics[height=8cm]{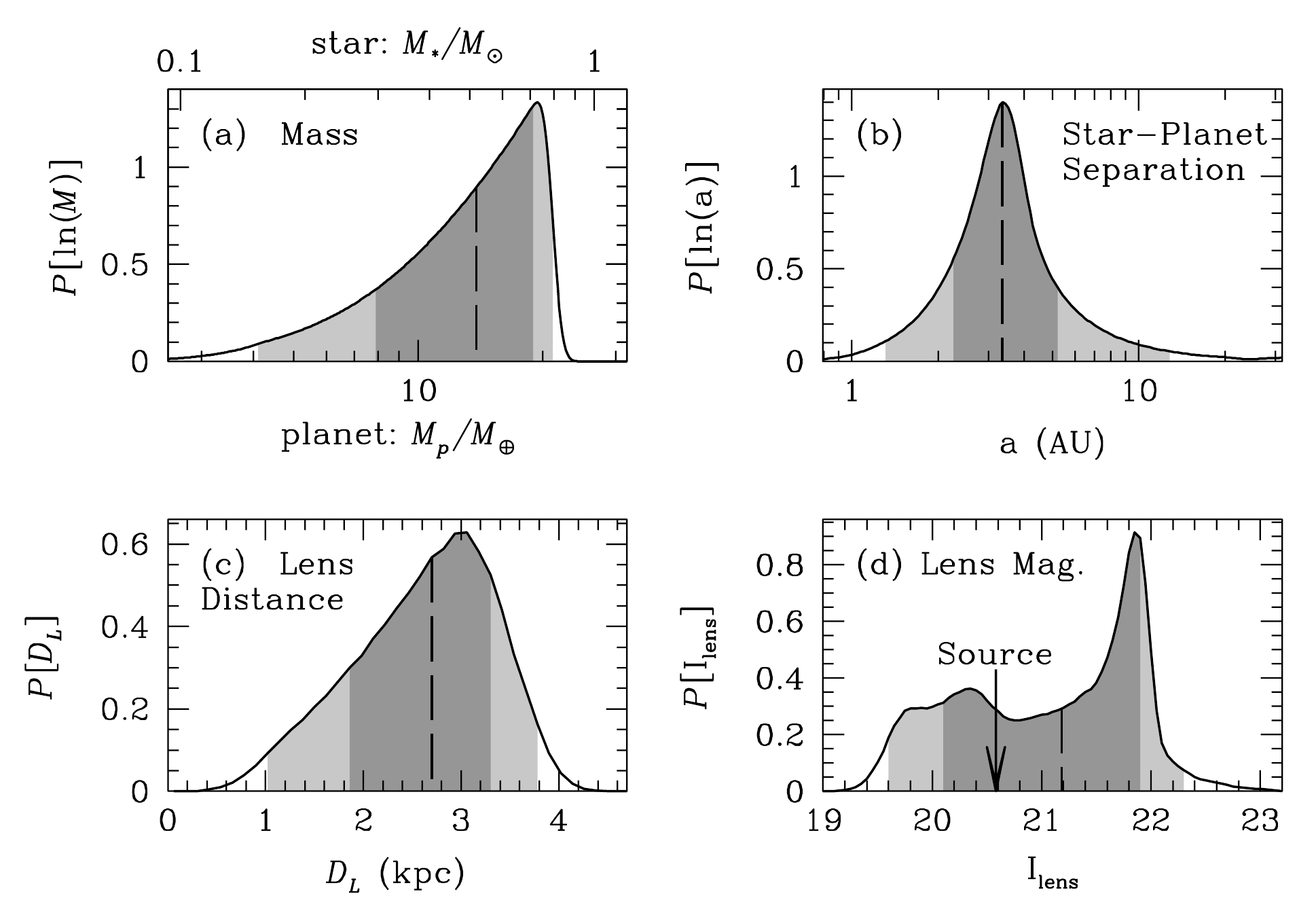}
\caption{OGLE-2005-BLG-169 lens property figure.
Bayesian probability densities for the properties of the planet
and its host star if it is a main sequence star. 
(a) The masses of the lens star and its planet ($M_\ast$
and $M_p$ respectively). 
(b) the separation, 
(c) their distance from the observer ($D_L$); 
and (d) the I-band brightness of the host star. 
The dashed vertical lines indicate the medians, and the shading
indicates the central 68.3\% and 95.4\% confidence intervals.  All
estimates follow from a Bayesian analysis assuming a standard model for the
disk and bulge population of the Milky Way, the stellar mass function of
Bennett \& Rhie (2002).}
\label{fig-ogle169_prop}       
\end{figure}

The analysis
of \citet{ogle169} indicated a super-Earth mass planet with
$M_p = 13 {+6\atop -8} \mearth$ orbiting a star of
$M_\ast \simeq 0.49 \msun$. Such a planet,
like OGLE-2005-BLG-390Lb,  would be
invisible to other planet detection methods. High magnification 
events also place tight constraints on the presence of additional
planets, and in the case of OGLE-2005-BLG-169, Jupiter-mass
planets can be exclude from the separation range 0.6-$18\,$AU
and Saturn-mass planets can be excluded from the range 1-$11\,$AU.

The precise masses of the host star and planet have not yet 
been determined because the host star has not been detected. 
Thus, the lens system properties can only
be determined by a Bayesian analysis, as was done for OGLE-2005-BLG-390Lb
in Fig.~\ref{fig-ogle390_prop}. This analysis uses
the parameters from the microlensing light curve, including
the Einstein radius crossing time of $t_E = 43\pm 1.3\,$days,
the source radius crossing time of $t_\ast = 0.019 \pm 0.01\,$days, and
the lens-source relative proper motion of 
$\mu_{\rm rel} = 8.4\pm 0.6\,{\rm mas/yr}$. The results of this analysis are
presented in Figure~\ref{fig-ogle169_prop}. These have
assumed a Han-Gould model for the Galactic bar
\citep{hangould-mpar}, a double-exponential disk with a
scale height of $325\,$pc, and a scale length of $3.5\,$kpc,
as well as other Galactic model parameters as described in
\citet{gest-sim}. Because this model is different from the
Galactic model used by \citet{ogle169}, the resulting parameters
differ slightly from their results. We find a lens
system distance of $D_L = 2.7{+0.6\atop -0.9}\,$kpc, a three
dimensional star-planet separation of $a = 3.3{+1.9\atop -0.9}\,$AU
and main sequence stellar and planetary masses of 
$M_\ast = 0.52 {+0.19\atop -0.22}\msun$ and 
$M_p = 14 {+5\atop -6}\mearth$. If we assume that white dwarfs
have an {\it a priori} probability to host planets that is equal to
that of main sequence stars (at the separations probed by 
microlensing), then there is a 35\% probability that the host 
star is a white dwarf. The possibility of a brown dwarf host star
is excluded by the light curve limits on the microlensing parallax
effect \citep{ogle169}.

Fig.~\ref{fig-ogle169_prop}(d) shows the probability distribution
of the $I$-band magnitude of the planetary host star compared to the
source star at $I = 20.58\pm 0.10$. The implied planetary host
star brightness distribution has a median and 1-$\sigma$ range of
$I_{\rm lens} = 21.9 {+0.7\atop -1.1}$, but the most interesting
feature of this figure is that the probability of a main sequence
lens fainter than $I = 23$ vanishes. This is because the 
mass-distance relation, eq.~\ref{eq-m_thetaE} ensures that the
lens star will be nearby and at least at bright as $I=23$, even if
it is at the bottom of the main sequence at $M_\ast = 0.08\msun$.
In fact, the microlensing parallax constraint from the light curve
yields a lower limit for the lens star mass of $M_\ast \simgt 0.14\msun$.
Thus, the planetary host star must be at least 16\% of the brightness of
the combined lens plus source star blended image, and this implies 
that it will be detectable if it is not a stellar remnant. Plus, the
relatively rapid relative proper motion, $\mu_{\rm rel} = 8.4\pm 0.6\,{\rm mas/yr}$,
of OGLE-2005-BLG-169L, implies that the lens-source separation
is already detectable with HST \citep{plan_char}, as discussed in
\S~\ref{sec-theta_E}.

One of the most interesting consequences of the discoveries of
OGLE-2005-BLG-390Lb and 169Lb is that super-earth planets
of $\sim 5$-$15\mearth$ are likely to be quite common. 
\citet{ogle169} combined these detections with null results
from very high magnification events \citep{abe-moa32,dong-ogle343}
plus samples of lower magnification events \citep{planet-limit,gaudi-planet-lim}
to solve for the fraction, $f_{\rm se}$, of stars with planets of mass
ratio $\sim 8\times 10^{-5}$ at the separations of 1.5-4$\,$AU,
where microlensing is most sensitive. They found that
the median and 90\% confidence level upper and lower
limits are $f_{\rm se} = 0.38{+0.31\atop -0.22}$, based on the two
planets discovered and the accumulated null results.
The 90\% c.l. lower limit is $f_{\rm se} \geq 16$\%. This is significantly
higher than the fraction of F, G, and K stars with Jupiter-mass
planets in this 1.5-4$\,$AU separation range. This fraction
of stars with Jupiters at this separation can be estimated from
\citet{butler-catalog} to be $f_{\rm J} \simeq 3$\%.
Thus, these cool, super-earth planets appear to represent the
most common type of exoplanet yet discovered.
This would seem to confirm the prediction of the core-accretion theory
that $\sim 10\mearth$ planets form much more frequently
than gas giants, like Jupiter \citep{ida_lin,laughlin}, although
this may not be incompatible with the disk instability theory
\citep{boss-supearth}.

\begin{figure}
\centering
\includegraphics[height=6.3cm]{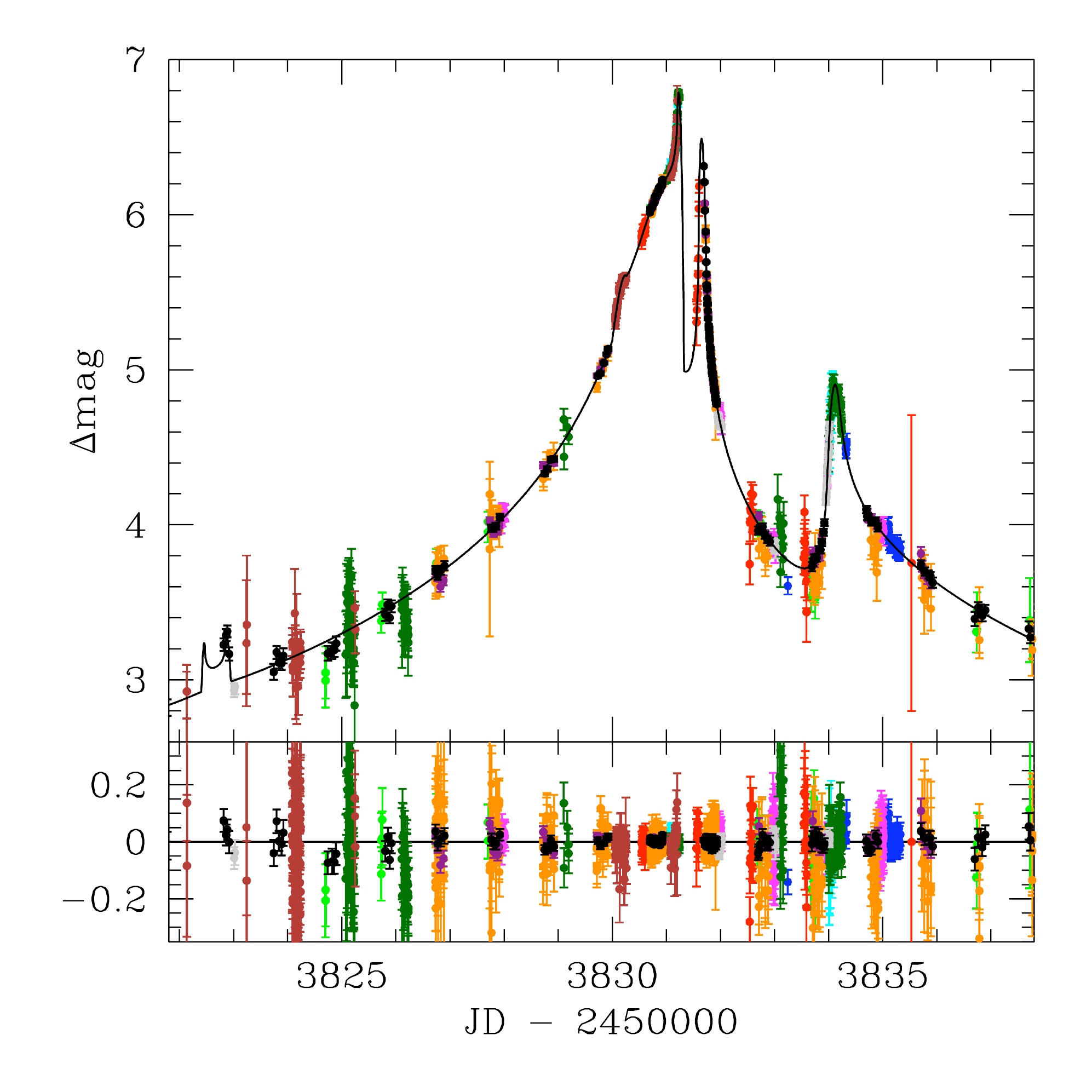}
\includegraphics[height=6.3cm]{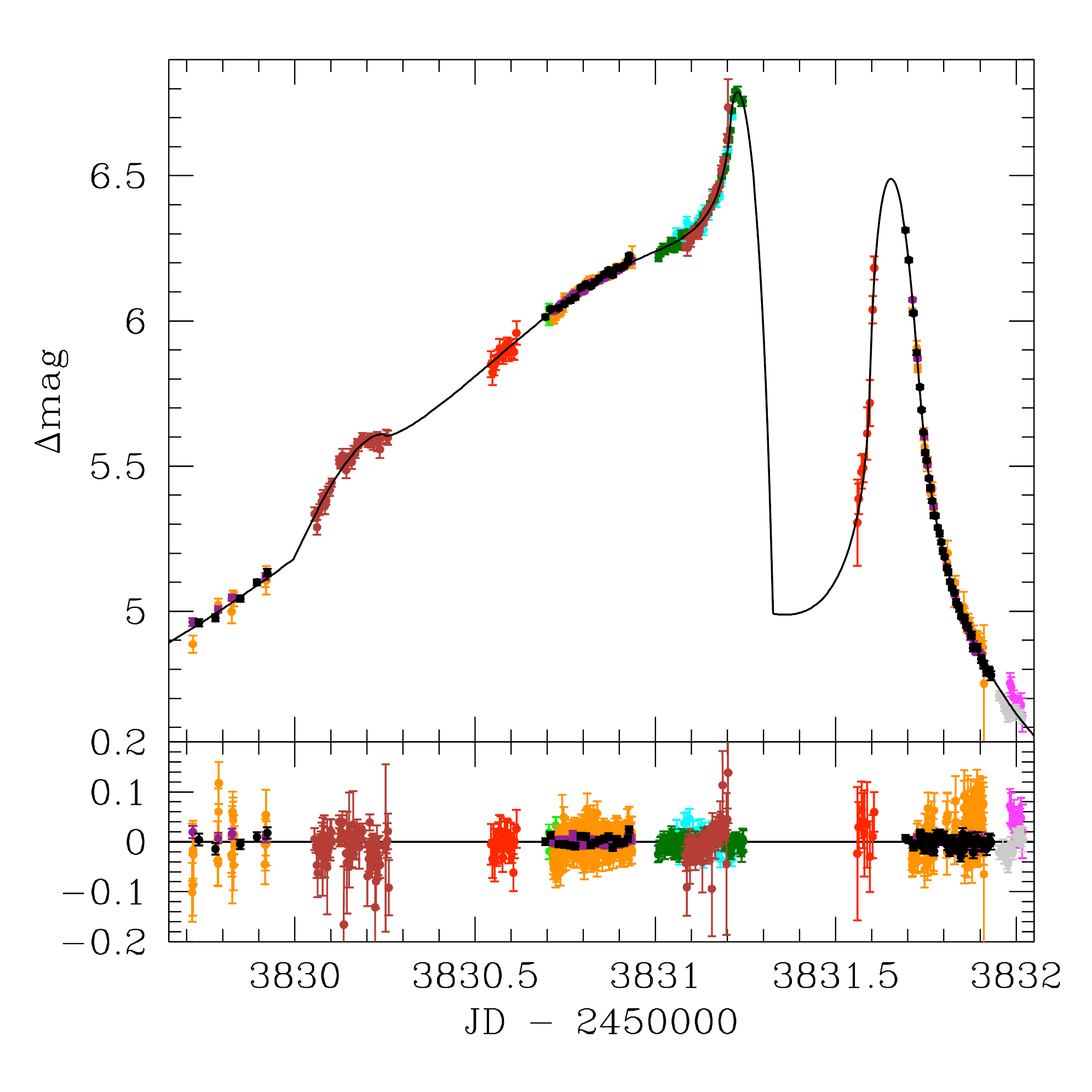}
%
%
\caption{Two views of the OGLE-2006-BLG-109 light curve, which is
the first multi-planet microlensing event with a planet of slightly less than
a Jupiter mass ($q = 1.35\times 10^{-3}$ at $\sim 2.7\,$AU and 
a planet of slightly more than a Saturn mass ($q = 4.9\times 10^{-4}$)
at $\sim 5.4\,$AU. The signal is dominated by the Saturn-mass
planet because it is close to the Einstein ring at $d = 1.4$, and there are 
two pairs of caustic crossing features (at $t = 3822.5$,  3822.9 and
$t = 3830.2$, 3831.2) and a cusp approach (at $t = 3834.1$) due to the
Saturn-mass planet. The Jupiter-mass planet planet is further
from the Einstein ring at $d = 0.63$, so its signal is limited to the
highest magnification part of the light curve and is responsible
for the cusp approach feature at $t = 3831.65$. Both planetary
orbital motion and microlensing parallax must be included to
obtain an acceptable model for this event.
}
\label{fig-ogle109_lc}       
\end{figure}

\begin{figure}
\centering
\includegraphics[height=5.4cm]{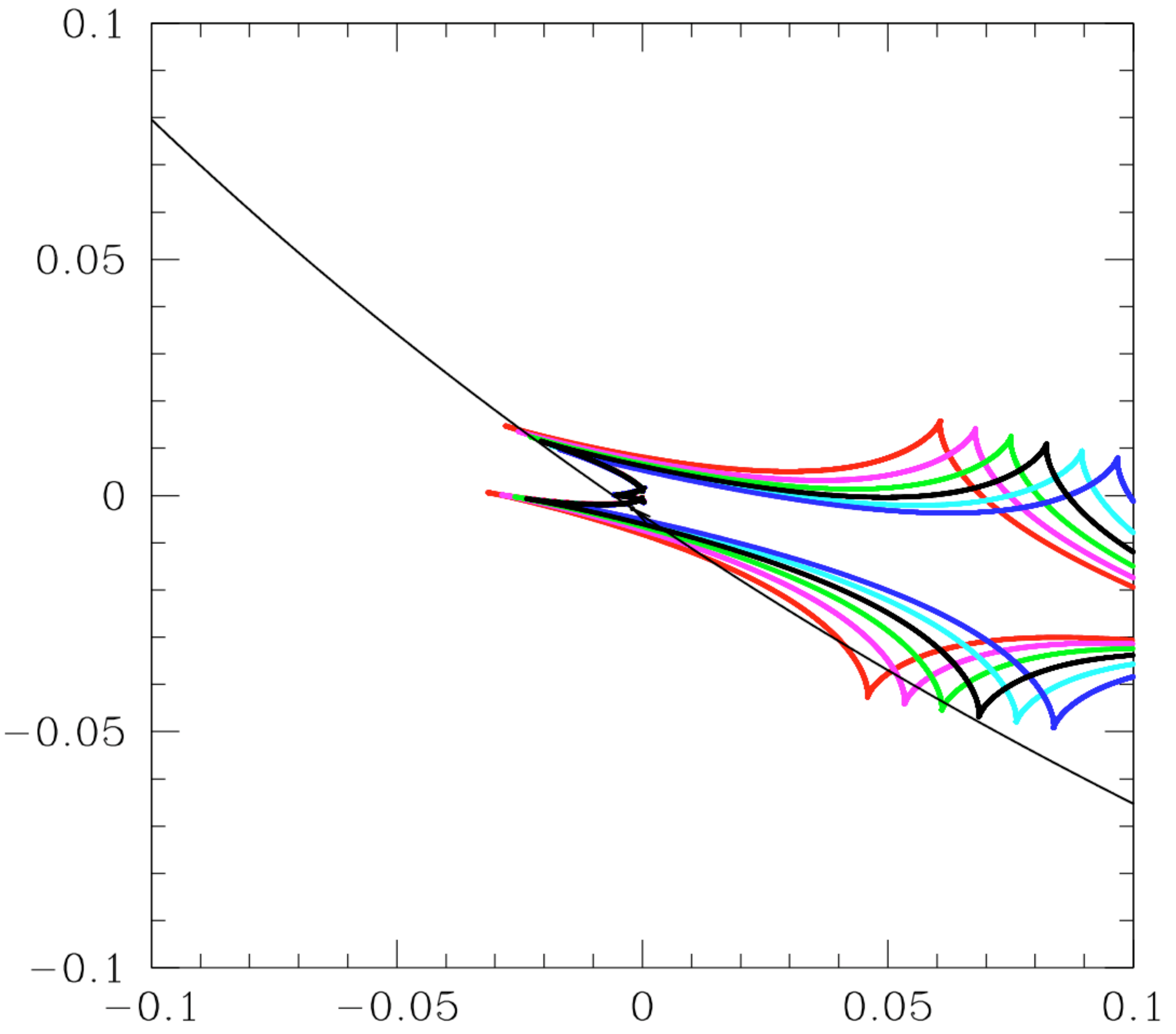}
\includegraphics[height=5.4cm]{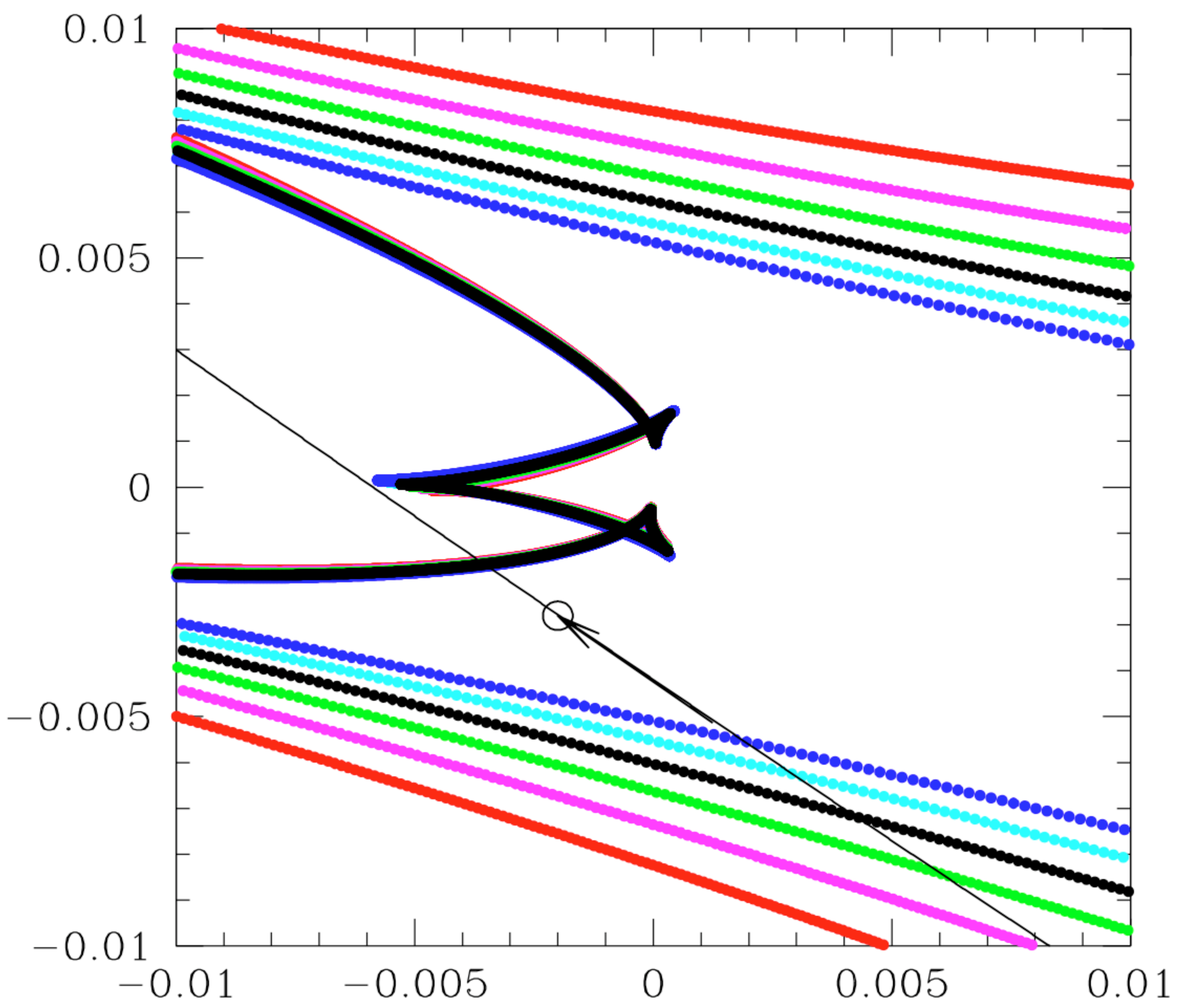}
\caption{The central caustic OGLE-2006-BLG-109 configuration
for is shown at 3-day intervals from $t = 3820$ (shortly before
the first caustic crossing) through $t = 3835$ (a day after the
final cusp approach). The time-order of the different color
caustic curves is {\it red, magenta, green, black, cyan, blue}.
The grey curve is the source trajectory, which is curved due
to the microlensing parallax effect (\ie the orbital motion
of the Earth) and the small circle that the source trajectory
in the left, close-up panel shows the source star radius.
}
\label{fig-ogle109_caus}       
\end{figure}

The final event that we will present is OGLE-2006-BLG-109, which is
much more complicated than the other events
(Gaudi \etal\ 2007; Bennett \etal\ 2007, both in preparation). 
The light curve for this event
is shown in Fig.~\ref{fig-ogle109_lc}, while the central caustic configuration
is shown in Fig.~\ref{fig-ogle109_caus}. This is the first microlensing
event with two detected planets, and it also shows clear signals of 
planetary orbital motion and microlensing parallax. These effects
are detectable because the Saturn-mass planet has a
projected separation that is close to the Einstein ring, which causes
its caustic to become quite extended. Its effects are visible for
11 days.

Another notable feature of OGLE-2006-BLG-109 is that the lens is
$> 5$ times brighter than the source. It is detectable (although not
completely resolved) in the best seeing (0.7") OGLE images and
is clearly visible in $K$ and $H$-band adaptive optics images
from the Keck telescope. As a result, there are two methods to
determine the lens star mass: the combination of the $\theta_E$
determination from the finite source effects and the microlensing
parallax effect yields the lens mass via eq.~\ref{eq-m_rep}, while
the lens star detection give the lens mass with the help of 
mass-luminosity relations, as discussed in \S~\ref{sec-theta_E}.
However, one complication is that there is some degeneracy
in the effects of microlensing parallax and the planetary orbital motion
on the microlensing light curve. On the other hand, the planetary
orbital motion parameters yield information about the orbits
that haven't been detected before in a microlensing event.
So, this event will yield much more information about the
OGLE-2006-BLG-109L planetary system than was anticipated
for any planetary microlensing event.

\section{Future Programs}
\label{sec-future}

Our experience with the existing microlensing planet search programs
provides indications of how the sensitivity of future microlensing surveys
can be improved. At present, the OGLE and MOA groups are each able to
independently discover more than 500 microlensing events per year.
There is a great overlap between the discoveries of these two groups,
but the total number of events discovered every year is probably
about 700. This is at least an order of magnitude larger than
the global follow-up groups can hope to follow. The follow-up groups
optimize their observations by focusing on high magnification events.
However, many of the shorter time scale high magnification events are
not recognized as such in time, and so a large fraction of the high
magnification events are not searched for planets.

The solution to this problem is to observe many microlensing events
in each image with a global network of very wide FOV telescopes that can observe
10 square degrees or more of the Galactic bulge at 15-20 intervals.
The new 1.8m MOA-II telescope \citep{moa2-tel} with a 2.2 square
degree FOV CCD camera \citep{moa2-cam} that began operation in
2006 is the first telescope that meets this requirement, and the OGLE
group plans to upgrade to a 1.4 square degree OGLE-4 camera in
time for the 2009 Galactic bulge observing season. WIth MOA-II
in New Zealand, and OGLE-IV in Chile, all that is needed is a
very wide-FOV microlensing survey telescope in Southern Africa.
A number of groups are pursuing funding for such a telescope.

\begin{figure}
\centering
\includegraphics[height=7cm]{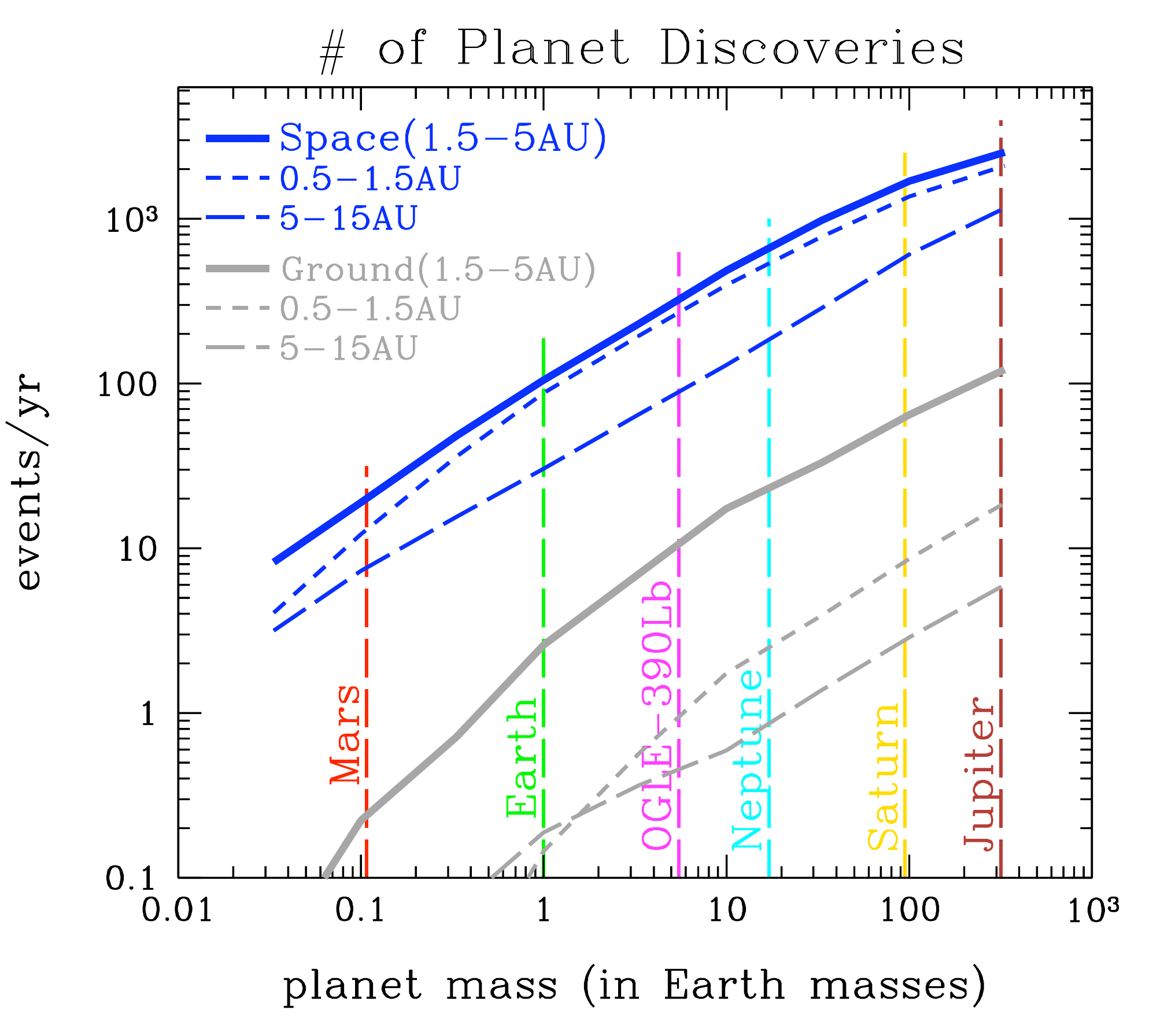}
\caption{
The number of planet detections expected per year as 
a function of planet mass is shown for proposed future space
and ground-based microlensing surveys under the assumption
of one planet per star in the indicated separation ranges.
The space-based survey has its most significant advantage
over the ground-based survey at separations smaller
(0.5-1.5$\,$AU) and larger (5-15$\,$AU) than the Einstein
radius, because a space-based survey is able to resolve
bulge main sequence stars and detect moderate
amplitude planetary signals when the magnification
due to the stellar lens is small.
\label{fig-fut_nplan}}
\end{figure}

Simulations of such a system have been performed by \citet{bennett-iap}
and Gaudi (2007, private communication), and estimates of the sensitivity
of the global network consisting of MOA-II, OGLE-IV and an OGLE-IV-like
system in South Africa are presented in Fig.~\ref{fig-m_v_sep} and
Fig.~\ref{fig-fut_nplan}. The improvement in sensitivity with such a network
in the mass vs. semi-major axis plane is shown in Fig.~\ref{fig-m_v_sep} 
with the light and dark red curves showing the sensitivity of the current
surveys and the future very wide-FOV network, respectively. This network
will extend the sensitivity of the microlensing method down to an Earth
mass at planet-star separations close to the Einstein ring radius ($\sim 2$-3$\,$AU).

The separation range where ground-based microlensing is most sensitive,
1-5$\,$AU corresponds to the vicinity of the so-called ``snow-line\rlap," which
is the region of the proto-planetary disk where it is cold enough for
water-ice to condense, The density of solids in the proto-planetary increases
by a factor of $\sim 4$ across the  ``snow-line\rlap," and as a result, the
core accretion theory predicts that this is where the most massive planets will form
\citep{ida_lin,laughlin,kennedy-searth}. According to this
theory, giant planets form just outside the ``snow line" where they
can accrete $\sim 10\mearth$ of rock and ice to form a core
that grows into a gas giant like Jupiter or Saturn via the run-away 
accretion of Hydrogen and Helium onto this core.
However, this theory also predicts that the Hydrogen and Helium
gas can easily be removed from the proto-planetary disk during the
millions of years that it takes to build the rock-ice core of a gas-giant.
Thus, if the core accretion theory is correct, rock-ice planets of 
$\sim 10\mearth$ that failed to grow into gas giants
should be quite common, although it is possible to form
such planets in the competing gravitational instability theory
\citep{boss-supearth}.

The number of planetary microlensing event detections expected per year
is shown in Fig.~\ref{fig-fut_nplan} assuming an average of one such planet
per star, with conservative assumptions regarding photometric precision.
The assumption of an average of one such planet per star is certainly
too optimistic for Jupiter mass planets \citep{gaudi-planet-lim,butler-catalog}, 
but it is closer to reality for super-earths, like OGLE-2005-BLG-390Lb and
OGLE-2005-BLG-169Lb \citep{ogle390,ogle169}. It could very well be accurate
for Earth-mass planets where the weaker two-body gravitational interactions
allow two planets to orbit in the separation range corresponding to the bins
in Fig.~\ref{fig-fut_nplan}. (Our own solar system is an example of this.)

Another future development that is already funded is a global network
of robotic telescopes dedicated to monitoring transient events like
planetary microlensing events, known as the Las Cumbres Global
Telescope Network \citep{lcogt}. Ideally, this network would routinely
observe high magnification microlensing events and planetary deviations
discovered in progress with an very high cadence, such as that 
provided by the MDM telescope for OGLE-2005-BLG-169 
(see Fig.~\ref{fig-ogle169_lc}. This would enable the very wide-FOV 
survey telescopes to maintain their normal sampling strategy so that
other planetary microlensing events would not be missed. This might 
add to the planet detection efficiency substantially, but such
a system is more difficult to model.

\subsection{The Ultimate Exoplanet Census: Space-Based Microlensing}
\label{sec-space}

The ultimate census for virtually all types of exoplanets would be a space-based
microlensing survey \citep{gest-sim,exoptf-MLspace}. Such a survey could 
provide a statistical census of exoplanets with masses $\geq 0.1\mearth$
and orbital separations ranging from $0.5\,$AU to $\infty$. This includes analogs 
to all the Solar SystemÕs planets except for Mercury, as well as most types of 
planets predicted by planet formation theories. This survey would determine
the frequency of planets around all types of stars except those with short lifetimes.
Close-in planets with separations $< 0.5\,$AU are invisible to a space-based 
microlensing survey, but these can be found by Kepler \citep{kepler}.
Other methods, including ground-based microlensing, cannot approach the 
comprehensive statistics on the mass and semi-major axis distribution of 
extrasolar planets that a space-based microlensing survey will provide.
Detailed simulations of a space-based microlensing survey
\citep{gest-sim} have been used to determine the sensitivity of
such a mission, and Figs.~\ref{fig-m_v_sep} and \ref{fig-fut_nplan} show
the sensitivity of the proposed Microlensing Planet Finder (MPF) mission
\citep{mpf-spie}. These figures also show that the sensitivity
of a ground-based microlensing survey to terrestrial planets is limited to
the vicinity of the Einstein radius at 2-3 AU. This is because ground-based
survey generally requires moderately high magnification $A \simgt 10$ 
in order to resolve the source star well enough to get the moderately
precise photometry that is needed to detect planets with the microlensing
method. A space-based microlensing survey would generally resolve the
source stars, so planets further from the Einstein radius can be detected
via their light curve perturbations at relatively low magnification 
from the lensing effect of the planetary host star. 

\begin{figure}
\centering
\includegraphics[height=4.4cm]{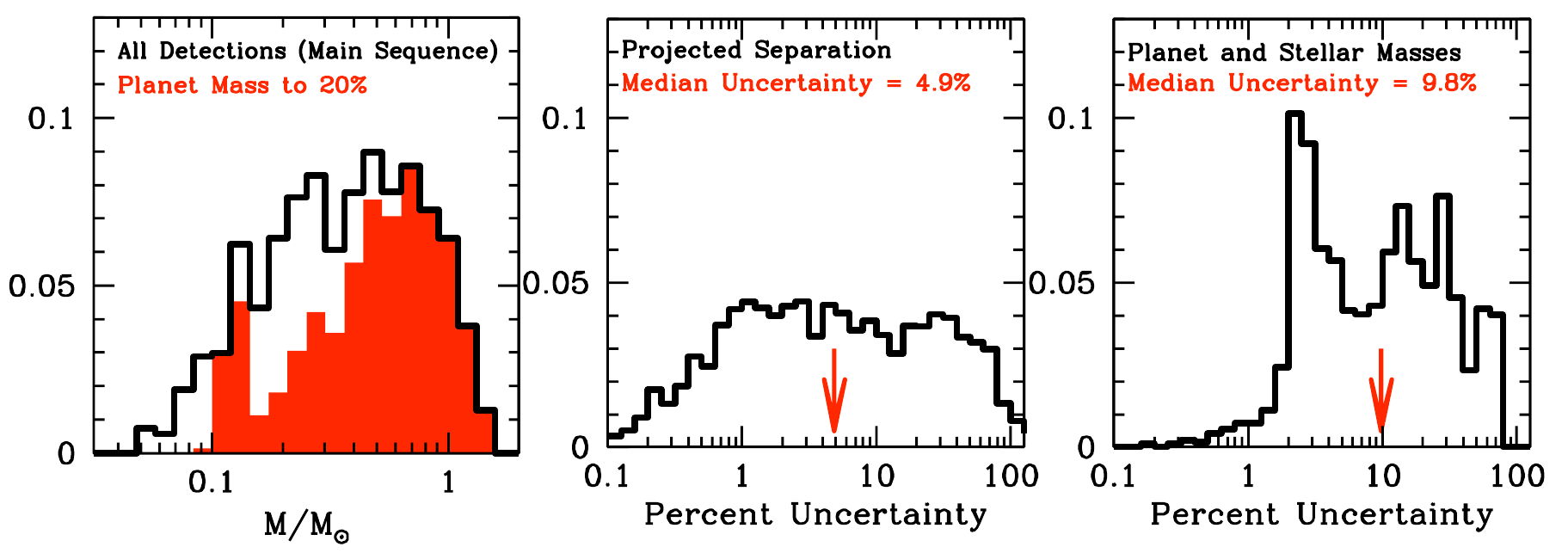}
\caption{
(a) The simulated distribution of stellar masses for stars with 
detected terrestrial planets. The red histogram indicates the
subset of this distribution for which the masses can be determined 
to better than 20\%.
(b) The distribution of uncertainties in the projected star-planet separation.
(c) The distribution of uncertainties in the star and planet masses. Note that
it is the two-dimensional projected separation that is presented here, and we
have not included the uncertainty in the separation along the line-of-sight
as was done in Figure~\ref{fig-ogle169_prop}.
\label{fig-hist_prop}}
\end{figure}

A space-based microlensing survey is also able to detect most
of the planetary host stars for most planetary microlensing events.
Using the methods described in
\S~\ref{sec-theta_E} and in more detail in \citet{plan_char}, 
this allows the determination of
the star and planet masses and separation in physical units.
This can be accomplished with HST observations for a small number 
of planetary microlensing events (Bennett et al. 2006), but only a
space-based survey can do this for hundreds or thousands of planetary 
microlensing events that future surveys would expect to discover.
Fig.~\ref{fig-ogle169_prop} shows the distribution of planetary host star 
masses and the predicted uncertainties in the masses and separation 
of the planets and their host stars \citep{plan_char} from 
simulations of the MPF mission. The host stars with masses determined
to better than 20\% are indicated by the red histogram in 
Fig.~\ref{fig-ogle169_prop}(a), and these are primarily the host stars 
that can be detected in MPF images.
Ground-based microlensing surveys also suffer significant losses in data 
coverage and quality due to poor weather and seeing. As a result, a 
significant fraction of the planetary deviations seen in a ground-based
microlensing survey will have poorly constrained planet parameters 
due to poor light curve coverage \citep{peale_gnd_v_space}. (These
poorly characterized detections are not included
in Fig.~\ref{fig-fut_nplan}, however.)

Proposed improvements to ground-based microlensing surveys can 
detect  Earth-mass planets in the vicinity of the ``snow-line\rlap,"
which is critical for the understanding of planet formation theories
\citep{gould-exoptf}.  But such a survey would have its sensitivity to 
Earth-like planets limited to a narrow range of semi-major axes, 
so it would not provide the complete picture of the frequency of 
exoplanets down to $0.1\mearth$ that a space-based microlensing 
survey would provide. Such a survey would probably not  
detect the planetary host stars for most of the events, and so it cannot 
provide the systematic data on the variation of exoplanet properties as a 
function of host star type that a space-based survey will provide.
 
A space-based microlensing survey, such as MPF, will provide a 
census of extrasolar planets that is complete (in a statistical sense) 
down to $0.1\mearth$ at orbital separations $\geq 0.5\,$AU, and when 
combined with the results of the Kepler mission a space-based microlensing 
survey will give a comprehensive picture of all types of extrasolar planets 
with masses down to well below an Earth mass. This complete coverage 
of planets at all separations can be used to calibrate the poorly understood 
theory of planetary migration. This fundamental exoplanet census data is 
needed to gain a comprehensive understanding of processes of planet 
formation and migration, and this understanding of planet formation is an 
important ingredient for the understanding of the requirements for habitable 
planets and the development of life on extrasolar planets
\citep{exoptf-MLspace}.

The basic requirements for a space-based microlensing survey are a 1-m class
wide field-of-view space telescope that can image the central Galactic bulge
in the near-IR or optical almost continuously for periods of at least several 
months at a time. This can be accomplished as a NASA Discovery mission, 
as the example of the MPF mission shows, but it could also be combined
with other programs that require an IR-optimized wide-FOV space telescope,
as long as a large fraction of the observing time is devoted to Galactic bulge
observations. As Fig.~\ref{fig-m_v_sep} shows, there is no other planned 
mission that can duplicate the science return of a space-based microlensing 
survey, and our knowledge of exoplanets and their formation will remain incomplete 
until such a mission is flown.

Thus, a space-based microlensing survey is likely to be the only way to gain 
a comprehensive understanding of the nature of planetary systems, which is 
needed to understand planet formation and habitability. The proposed 
Microlensing Planet Finder (MPF) mission is an example of a space-based 
microlensing survey that can accomplish these objectives with proven 
technology and a cost that fits comfortably under the NASA Discovery Program cost cap.

%



\printindex
\end{document}